\documentclass{aa}

\usepackage{graphicx}
\usepackage{siunitx}
\usepackage{txfonts}
\usepackage{placeins}
\usepackage[dvipsnames]{xcolor}
\usepackage{color}
\usepackage[compact]{titlesec}
\titlespacing{\section}{0pt}{*4}{*2}
\titlespacing{\subsection}{0pt}{*4}{*2}
\titlespacing{\subsubsection}{0pt}{*4}{*2}
\usepackage[colorlinks=true, linkcolor=blue, citecolor=blue, urlcolor=blue]{hyperref}
\definecolor{mhi}{rgb}{0.6,0.0,0.6}

\begin{document} 

\newcommand{\missingref}{\textcolor{red}{(Reference)}}

   \title{Intracluster globular clusters as tracers of the mass assembly of the Hydra\,I galaxy cluster}

   \author{Felipe S. Lohmann
          \inst{1}
          \and
          Magda Arnaboldi
          \inst{1}
          \and
          Michael Hilker
          \inst{1}
          \and
          Andreas Burkert
          \inst{2}
          \and
          Marilena Spavone
          \inst{3}
          \and
          Ortwin Gerhard
          \inst{4}
          \and
          Marina Rejkuba
          \inst{1}
          \and
          Marco Mirabile
          \inst{1, 5, 6}
          \and
          Michele Cantiello
          \inst{5}
          \and
          Enrichetta Iodice
          \inst{3}
}

   \institute{European Southern Observatory, Karl-Schwarzschild-Stra\ss e 2, 85748, Garching, Germany\\
              \email{fschmidt@eso.org}
    \and
    Universitätssternwarte der Ludwig-Maximilians Universität, Scheinerstra\ss e 1, 81679 München, Germany 
    \and
    INAF -- Astronomical Observatory of Capodimonte, Salita Moiariello 16, I-80131, Naples, Italy
    \and
    Max-Planck-Institut für Extraterrestrische Physik, Giessenbachstra\ss e, 85748 Garching, Germany
    \and
    INAF -- Astronomical Observatory of Abruzzo, Via Maggini, 64100 Teramo, Italy
    \and
    Gran Sasso Science Institute, viale Francesco Crispi 7, I-67100 L'Aquila, Italy
    }

   \date{Received MM DD, YY; accepted MM DD, YY}

\date{Received XXX; accepted YYY}

   \abstract
   {
   In galaxy clusters, the hierarchical model of galaxy assembly predicts the formation of stellar substructures and intracluster light (ICL), a diffuse component consisting of stars that are not gravitationally bound to any single galaxy but instead follow the global gravitational potential of the cluster. 
   These features encode the details of the cluster's assembly history.
   However, observations are challenging due to their faint surface brightness, so independent tracers such as intracluster planetary nebulae and globular clusters (GCs) can provide valuable insight.
   For this work, we used deep VLT/FORS $V$- and $I$-band imaging to study the GC population in the Hydra\,I galaxy cluster, a rich environment of galaxies that is located at a distance of 45.7 Mpc.
   Our photometric sample of GC candidates was constructed from the $VI$ colour-magnitude diagram, where point sources with a similar colour as confirmed GCs were selected. 
   Dividing our GC sample in two colour groups, we show a striking difference between the two populations: while red GCs tend to be clustered around Hydra’s massive galaxies (mainly NGC\,3311 and NGC\,3309), blue GCs are more extended and spatially coincide with the peak of the cluster’s X-ray emitting gas.
   The GCs around the central galaxies also have different spatial distributions according to their stellar population properties. 
   Young metal-rich GCs are more extended and may be associated with ram-pressure tails, whereas old metal-poor GCs are more concentrated and could be related to disrupted dwarfs. The red, old, and metal-rich GCs are likely associated with the central massive galaxies.
   Comparing the GC number density profiles to the surface brightness profile of NGC\,3311, we find that the red GCs closely follow the galaxy’s light, while the blue population significantly deviates from it and traces the global gravitational potential of the cluster. 
   This result is also evidenced by the specific frequency of blue GCs, which is {$\sim$5 times} larger in the ICL-dominated outskirts when compared to the inner parts of the cluster and to the red population.
   Finally, we introduce a novel method to constrain the evolution of the galaxy luminosity function of the cluster from GC specific frequencies and colour distributions.
   This method results in a past Schechter slope {of $\alpha=-1.81_{-0.16}^{+0.16}$ for the faint end compared to $\alpha=-1.41_{-0.05}^{+0.08}$ in the present day}, which is consistent with measurements at high redshift and with cosmological simulations.
   }
 

   \maketitle

\section{Introduction}

In the current cosmological framework, galaxy clusters are the largest gravitationally bound structures in the universe. Their assembly occurs through a hierarchical process, involving the continuous merging of smaller structures such as massive galaxies and galaxy groups. This process is still happening today \citep{Kravtsov2012}.

As a result of this hierarchical assembly, the dense regions within galaxy clusters host a faint stellar component known as intracluster light (ICL). 
This diffuse light consists of stars that are not gravitationally bound to any single galaxy but instead follow the global gravitational potential of the cluster, with the first example of a kinematically confirmed sample of intracluster stars found in the Virgo cluster \citep{Arnaboldi96}. 
The kinematics of ICL provides key insights into the cluster's recent assembly history \citep[see review by][]{Arnaboldi2022}, while the ICL mass fraction correlates with the cluster's assembly time and dynamical history \citep{MontenegroTaborda25}.

Several mechanisms contribute to the formation of ICL \citep[see review by][]{Montes2022}, with the most significant being the stripping of stars from satellite galaxies and their subsequent mergers. 
In particular for $z=1$, mergers with the brightest cluster galaxy (BCG) dominate the ICL formation at small cluster radii, while tidal stripping dominates in the outskirts \citep{Murante07}. 
These processes contribute not only to the intracluster population of stars, but also to that of star clusters \citep{Muzzio86, White87, Hilker99}.
Each formation mechanism leaves distinct signatures on the properties of ICL stars, making ICL a key source of information on the history of galaxy interactions, dark matter distribution within clusters, and the dynamics of cluster evolution.

However, due to its faint nature ($\mu_g\gtrsim\,$\SI{26}{mag/arcsec^2}), ICL is challenging to study by direct imaging. 
In the last decade, deep imaging surveys have provided excellent results in estimating ICL \citep{Mihos17, Spavone24}.
Alternatively, individual globular clusters (GCs) and planetary nebulae (PNe) are detectable in nearby galaxy clusters and can serve as luminous point sources that trace the faint halos of galaxies and ICL, thus providing crucial insights into the assembly history of clusters \citep[e.g.][]{Arnaboldi96, Hilker99, Schuberth2008, Longobardi15a, AlamoMartinez2017, Harris20, Hartke2022, Chaturvedi2022}.
Recent JWST observations have additionally confirmed the link between GCs and ICL in clusters at intermediate redshifts \citep{Diego24, Martis2024}.
In particular, blue GCs are known to trace the X-ray-emitting intracluster medium \citep{Forbes2012, Cantiello2018} and thus share the same gravitational potential, suggesting a direct connection to the ICL.

The ratio of GCs to stars is not constant for all galaxies; rather, it depends on their properties \citep[e.g.][and references therein]{Lamers2017}.
The specific frequency of GCs, $S_N$, is a measure of the number of GCs per unit of galactic luminosity \citep{harris81}, and it measures the efficiency of galaxies in producing (and retaining) GCs.
It is defined as 
\begin{equation}
\label{eq:SN}
S_N \equiv N_{GC} 10^{0.4 (M_V + 15)}\ ,     
\end{equation}

where $N_{GC}$ is the number of GCs and $M_V$ is the $V$-band absolute magnitude of the galaxy.
This quantity has a known dependence with galaxy magnitude and metallicity \citep{Georgiev2010, Lamers2017}, with high $S_N$ measured both in dwarf ($M_V \gtrsim$ \SI{-18}{mag}) and giant ($M_V \sim$ \SI{-22}{mag}) galaxies, and low $S_N$ measured in intermediate mass ($M_V \sim$ \SI{-20}{mag}) ellipticals.
Similarly, the GC colour distribution is not independent of galaxy magnitude: bright galaxies show a distinctive bimodality in GC colour, with a red (metal-rich) and a blue (metal-poor) population, while fainter galaxies are dominated by the blue GC population \citep{Peng2006}.
Therefore, the specific frequency and colour distribution of GCs measured in the intracluster space may hint at the galaxies that were destroyed by tidal interactions and mergers to build the intracluster population of stars at that location in the cluster. For an example of using intracluster GC (ICGC) properties to reconstruct the assembly history of ICL in clusters, see the simulations by \citet{Ahvazi24}.

The galaxy luminosity function (LF) quantifies the number of galaxies per luminosity interval and has been used to constrain models of galaxy evolution in the cold dark matter (CDM) framework \citep{Benson03, Samui07}.
It is well described with a \cite{Schechter1976} function, given as

\begin{equation}
    \phi(M) = \frac{\mathrm{ln} 10 }{2.5} \phi^\star \left[10^{0.4(M^\star - M)}\right]^{\alpha+1} \mathrm{exp} \left[ -10^{0.4(M^\star - M)}\right],\,
\end{equation}

where $\phi^\star$ is a normalisation parameter, $M^\star$ is a characteristic magnitude and $\alpha$ is the faint-end slope.
For dark matter halos, the mass function has a theoretically expected faint-end slope close to $\alpha=-1.8$ \citep{PressSchechter74}, which is steeper than that observed in the local universe \citep{Trentham02}.
An evolving luminosity function finds observational support \citep{McLeod21, Ilbert05}, with a steeper faint-end slope at higher redshifts, approaching the CDM prediction.

As ICGCs are expected to arise mostly from the gravitational removal of GCs from satellite galaxies that did not survive until today
\citep{Ramos2020}, we can use the GC populations in galaxy clusters to learn about the evolution of the galaxy LF.
Most GCs are old and thus were formed back when the galaxy LF may have been steeper. 
Using the relations that connect the GC specific frequency and colour distribution with galaxy magnitude (\citealt{Georgiev2010} and \citealt{Peng2006}, respectively), we attempt to constrain the ancient galaxy population that dynamically evolved to assemble the galaxies of the galaxy cluster (and its intracluster populations) as we observe today.
Our testbed is the Hydra\,I galaxy cluster.

The Hydra\,I cluster is a rich environment of galaxies that is located
relatively nearby, at a distance of 45.7 Mpc \citep{Blakeslee01}.
The cluster is still in a dynamically evolving state, as evidenced by spectroscopic studies around the cluster centre \citep{Ventimiglia11, Coccato11} and recent surveys targeting low surface brightness features \citep{Capaccioli2015, Spavone24}, which found signatures of cluster galaxies interactions in the intracluster space.
Around the central galaxy NGC\,3311, \cite{Arnaboldi2012} and \cite{Barbosa2018} have shown that there exists an extended stellar envelope that is offset from the central galaxy both spatially and kinematically, and thus is associated with the global cluster gravitational potential. 
Furthermore, this faint, displaced component was shown to coincide with the peak of X-ray emission obtained by \cite{Hayakawa06}, who also found a high metallicity blob in the same region. 
Finally, \cite{Hilker2018} measured a steeply rising velocity dispersion profile for the stellar light around NGC\,3311, which can partly be explained by overlapping kinematic substructures, further underpinning the recent active assembly history of the cluster core.

The ICL in Hydra\,I was studied in \cite{Spavone24}.
They reported that the ICL is mostly concentrated around the cluster core and in a group of galaxies to the north of the cluster centre.
Additionally, the authors measured the surface brightness (SB) profile of the BCG+ICL component out to $0.4\,R_\textrm{vir}$ and found that an extended exponential component dominates the light for galactocentric distances larger than $\sim$ \SI{60}{kpc}.
These measurements, together with the extended spatial distribution of GCs, will allow us to measure the specific frequency of the ICL, which will then be used to constrain the number of destroyed galaxies necessary for its formation.

Previous works have studied the GC population in Hydra\,I.
Using spectra from the VIMOS instrument at the VLT, \cite{Misgeld2011} identified GCs and UCDs (ultra-compact dwarfs) around the centre of Hydra\,I and spectroscopically confirmed their association to the cluster.
They found a higher velocity dispersion for the brighter objects and suggested that these are stripped nuclei of dwarf galaxies.
Using the same data, \cite{Richtler2011} concluded that the kinematics of NGC\,3311 is better reproduced by a cored dark matter halo rather than a cuspy one. 
Finally, using MUSE data, \cite{Grasser24} reported that the GC population close to NGC\,3311 is old, metal-rich ($[M/H] \simeq 0.19 \pm 0.03 $ ) and $\alpha$-enhanced ($[\alpha/Fe] \simeq 0.12 \pm 0.03$).

In this paper, we study the ICL in the Hydra\,I cluster as traced by its ICGCs, particularly the blue GC population. 
We constrain the properties of the ICL at different clustercentric distances and investigate whether the ICL stars are predominantly originating from major mergers followed by violent relaxation \citep{Murante07}, or from the tidal disruption of dwarf galaxies \citep{Barai09}.
This problem is addressed using the properties of the GC spatial distributions, number densities and specific frequencies (Sec. \ref{sec:spatial_dist}).
We describe a novel method to study the evolution of the galaxy luminosity function by constraining the galaxies that were in place before the build-up of the ICL, using both the specific frequency and the colour distribution of intracluster GCs (Sec. \ref{sec:LF_evo}).
We discuss our results in Sec. \ref{sec:discussion} and summarise them in Sec. \ref{sec:conclusion}.

\section{Data}

\begin{figure}
   \centering
   \includegraphics[width=\columnwidth]{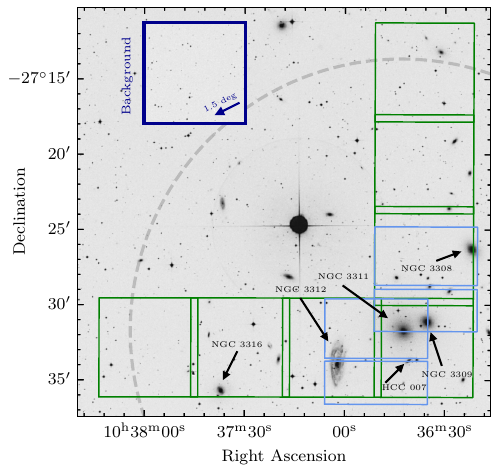}
   \caption{Optical image of the centre of the Hydra\,I cluster obtained from the Digitized Sky Survey. The green squares show the region covered by the seven $6.8' \times 6.8'$ FORS1 pointings from which we obtained the $V$ and $I$ photometry for our GC candidates, while the blue squares show the regions with $U$-band coverage.
   The dark blue inset shows the background field located outside the cluster in the direction indicated by the arrow.
   The dashed grey circle corresponds to a radius of \SI{0.15}{R_{vir}} = \SI{240}{kpc}.}
   \label{fig:photometry_pointings}
\end{figure}

The imaging data used in this work was acquired with the Very Large Telescope using the FORS1 instrument at UT1 (Observing Programmes 65.N–0459(A), PI: M. Hilker; 80.A-0647(B), PI: M. Hilker; 82.A-0894(A, B), PI: M. Hilker) in imaging mode. 
The observations were carried out in the $V$, $I$ and $U$ bands with a maximum seeing of $\leq$ \SI{0.6}{arcsec} for the $V$ and $I$ bands, and $\leq$ \SI{0.8}{arcsec} for the $U$ band, and consist of nine $6.8' \times 6.8'$ fields at a pixel scale of \SI{0.2}{arcsec/pixel}.
The $V$ and $I$ observations consist of seven fields arranged in an L shape centred on the central region of the Hydra\,I cluster, while the $U$-band observations consist of two fields covering the four brightest galaxies of the cluster.
For each field we acquired 3$\times$\SI{480}{s} exposures in the $V$ band, 9$\times$\SI{330}{s} exposures in the $I$ band, and (6$\times$\SI{1250}{s}+1$\times$\SI{300}{s}) exposures in the $U$ band.
Additionally, one background frame with the same exposure in each band was taken in an empty region $\sim$ \SI{1.5}{degrees} ($\sim$ \SI{1.2}{Mpc}) away from the cluster centre to perform a statistical subtraction of background contaminants from our GC sample \citep{Cantiello2015, Mirabile24}.
Figure \ref{fig:photometry_pointings} shows the seven $VI$ and the two $U$ fields, as well as the background field.

{The data reduction follows the same steps as carried out by \cite{MieskeHilker03} for the Centaurus cluster, including bias subtraction, flat-fielding, image co-addition, and cosmic ray removal. 
The photometric calibration was done through dedicated standard star observations close in time with respect to science. 
The PSF was built using a selected list of stars with dedicated DAOPHOT routines in IRAF. 
The PSF was modelled using a \cite{Moffat69} distribution with a coefficient $\beta = 2.5$, and the sky was estimated locally within a ring around each source. To calibrate the PSF magnitudes we performed aperture photometry for the selected stars used to derive the PSF model.}
Finally, foreground galactic extinctions for NGC\,3311 were based on the reddening measurements from \cite{Schlafly2011}, and correspond to $A_V = 0.217$, $A_I = 0.119$, $A_U = 0.343$ for the $V$, $I$ and $U$ bands, respectively, and we assume a distance modulus for Hydra\,I of 33.3 mag \citep{Blakeslee01}, corresponding to a distance of 45.7 Mpc and implying an angular scale of \SI{13.3}{kpc/arcmin}.
For NGC\,3311 ($\alpha_{J2000}=\,$10:36:42.80, $\delta_{J2000}=\,$-27:31:41.23), we assume an effective radius $R_e=\,$\SI{8.4}{kpc} \citep{Arnaboldi2012}.

\section{Methods}

\subsection{Point source identification and GC selection}
\label{sec:sample_selection}

At the distance of the Hydra\,I cluster, GCs are unresolved point sources.
For reference, Milky Way GCs have median half-light radius of $\sim$ \SI{3}{pc} \citep{Harris96}.
{The detection and selection of point sources in the reduced images were done after modelling and subtracting the galaxy light of NGC 3311 and NGC 3309. The galaxy light was modelled iteratively using IRAF’s task ellipse.}
Then, the \textsc{rmedian} task was used to subtract additional extended sources from the images.
To select point sources from the galaxy-subtracted images, we used two parameters from the \textsc{daophot} PSF photometry, which was performed in the galaxy-subtracted images.
The first parameter is sharpness, which roughly quantifies the difference between the square of the width of an object and the square of the width of the PSF.
Single stars have a sharpness close to zero, whereas extended objects have larger positive or negative values.
The second parameter used is the goodness of PSF fit statistic $\chi$, that quantifies the pixel-to-pixel scatter from the fitting residuals.

We therefore seek objects with small absolute values of sharpness and $\chi$. We defined the limits in these parameters as the exponential curves shown in the top and middle panels of Fig. \ref{fig:point_source_selection_vi}. 
These are constructed rather arbitrarily, such that fainter sources may have larger sharpness but the lowest $\chi$.
Additionally, we restrict our selection to objects with photometric errors below \SI{0.2}{mag}.
{In the central field covering NGC\,3311 and NGC\,3309, our PSF photometry resulted in significantly lower $\chi$ values, and thus we used more restrictive selection curves (see insets in Fig. \ref{fig:point_source_selection_vi}).}
The sources falling within our defined limits in sharpness, $\chi$, and photometric errors simultaneously, both in the $V$ and $I$ bands, make up our sample of point sources, which consists of {9655} objects. These are plotted as black points in Fig. \ref{fig:point_source_selection_vi}.

The next step was to select GC candidates from these point sources.
We used spectroscopically confirmed Hydra\,I GCs from \cite{Misgeld2011} to select point sources with similar $(V-I)_0$ colour.
We cross-matched the spectroscopic and photometric catalogues for point sources with $V<\,$ \SI{24}{mag}, as this is the magnitude limit reached by the spectroscopic survey.
{The 54 matched}, confirmed GCs are then used to constrain the $(V-I)_0$ colour and select GC candidates from our photometric data.

\begin{figure}
   \centering
   \includegraphics[width=\columnwidth]{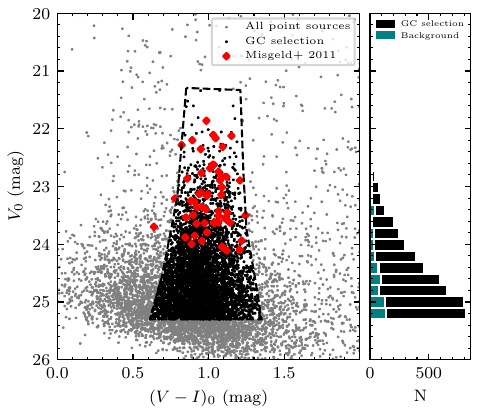}
   \caption{GC candidates' selection using the $VI$ colour-magnitude diagram. The red diamonds are point sources from \cite{Misgeld2011} that are spectroscopically confirmed to be associated with Hydra\,I. These objects reveal the locus of GCs in the colour-magnitude space, and we use them to select our sample of GC candidates, shown as the black points delimited by the dashed polygon.
   The right panel shows the $V$-band luminosity function of our selected candidates and of the objects in the background field, normalised to the area covered by the seven FORS fields.}
              \label{cmd_vi}
    \end{figure}

Figure \ref{cmd_vi} shows the $VI$ colour-magnitude diagram for our sample of point sources, as well as the spectroscopically confirmed GCs that were successfully matched.
Most of them occupy a well defined region in this colour-magnitude space, with $(V-I)_0$ colour {ranging from \SI{0.8}{mag} to about \SI{1.3}{mag}}, which is then used as constraint for the colour selection of our photometric GC candidates. 
Since our observations are deep, we extend the $V_0$ magnitude {limit down to $\sim$\SI{25.3}{mag}}, where incompleteness starts to affect our sample.
The completeness correction to our GC number counts is discussed in Sec. \ref{sec:radial_profiles}.
Our final selection of GC candidates is composed of {4886 objects} and is shown as the black points in Fig.~\ref{cmd_vi}.

The same point source and GC selection steps were carried out for the background field to obtain a statistical number of contaminants in our GC sample.
In the background frame, {112 objects} were selected following our method.
The right panel of Fig. \ref{cmd_vi} shows the $V$-band luminosity function of our selected GC candidates, as well as the expected numbers for background sources, rescaled to the area covered by the science frames.
The background contamination stays below $\lesssim 20\%$ for almost all magnitude bins.
Here, we assume there is a negligible number of ICGCs in the background field, meaning that all the selected objects are contaminants.
We make this assumption given that the field is $0.75 R_\textrm{vir}$ away from the centre and has no Hydra I galaxies nearby.

While $V$ and $I$ can be used to select GC candidates, they are not enough to constrain their stellar population properties. 
For example, GCs with a blue $V-I$ colour may either be young and metal-rich, or they could be old and metal-poor.
These properties cannot be disentangled from the $VI$ data alone, and we thus resort to the $U$-band photometry in the centre of Hydra\,I.
Although spatially limited, this data covers most of the objects with clustercentric distance $d \leq$ \SI{6}{arcmin}, except the non-overlapping sector to the southwest of NGC\,3311 for which only $VI$ data is available (see Fig. \ref{fig:photometry_pointings}).
We select point sources in the $UVI$ fields in a similar way as for the $VI$ fields, with cuts in sharpness, $\chi$, and photometric error in the three bands (see Fig. \ref{fig:uvi_point_sources}). 
This sample of point sources is then matched to the $VI$ GC candidates, which results in {1733 candidates} with $UVI$ photometry.
Using the $U$-band magnitudes, we can partially disentangle the age-metallicity degeneracy of GCs in a colour-colour plane \citep[e.g.][]{Cantiello2018b}.

\begin{figure*}
    \centering
    \sidecaption
    \includegraphics[width=\textwidth]{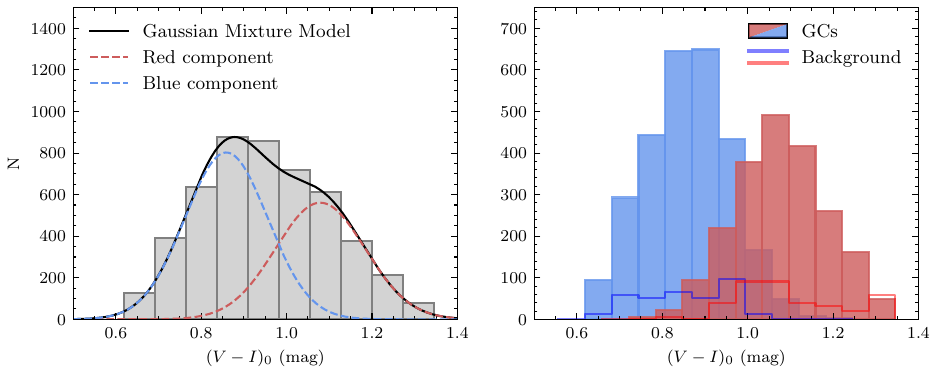}
    \caption{Left: Two-component Gaussian mixture model that best describes the $(V-I)_0$ colour distribution of our GC candidates (grey histogram). The individual components are shown as the blue and red dashed lines.
    Right: Red and blue GC subsamples obtained by assigning each GC to one Gaussian component. The expected numbers of background contaminants are shown as the step histograms.}
    \label{fig:gmm}
\end{figure*}

Globular cluster populations in bright and massive galaxies are usually bimodal in colour mainly due to their metallicity: blue GCs tend to be metal-poor, while red GCs are metal-rich \citep[e.g.][]{Muratov10}.
We separate our GC sample in $(V-I)_0$ colour in order to obtain red (rGCs) and blue (bGCs) GC subsamples.
To do this, we fitted a two-component Gaussian mixture model to the $(V-I)_0$ colour distribution of our sample, shown in the left panel of Fig. \ref{fig:gmm}.
We verify that a bimodal model provides the best description of the data given its lowest Bayesian inference criterion when compared with mixture models with different numbers of components.
Then, the GCs are split into red and blue in a statistical way. 
By treating each Gaussian component as a probability distribution, we assign each GC to one group according to their probabilities.
The objects selected in our background field are also assigned to red and blue background objects using the same Gaussian components.
The right panel shows the colour distributions of the red and blue GC subsamples, as well as those of background objects rescaled to the area of our field of view.

\subsection{GC radial profiles}
\label{sec:radial_profiles}

To characterise the overall spatial distribution of GCs and stars, we compare the number density profile of GCs with the stellar surface brightness of NGC\,3311 from \cite{Spavone24}. We calculated the number density profile for our GCs in annular bins centred on NGC\,3311.
The annuli are not constant in width, but instead are constructed such that each bin contains a fixed number of GCs, resulting in wider annuli at larger clustercentric distances. 
For a profile with 20 bins, {we obtained $\sim 240$ candidates} in each bin, with the uncertainties in the GC counts taken as Poissonian.
Additionally, we masked out the GCs located around the bright galaxies NGC\,3309 and NGC\,3316 using circular masks of $0.9$ arcmin in radius, which corresponds to $\sim$0.7 $R_e$ and $\sim$4.0 $R_e$, respectively.
The small mask (in terms of effective radii) chosen for NGC\,3309 is motivated by the fact that NGC\,3311 lies only $\sim$ \SI{1.6}{arcmin} away, whose GCs must be included in the number density profile.
The same procedure was carried out for red and blue GCs separately.

{We corrected our GC sample for incompleteness using injection–recovery experiments with artificial stars. 
Artificial sources were added to the science images over a range of magnitudes and galactocentric radii and recovered using the same detection and selection procedures applied to the real data. 
From these tests, we derived completeness curves in different radial bins, which are shown in Appendix \ref{appendix:completeness}.
Considering our GC selection limiting magnitude of \SI{25.3}{mag}, the curves show that bGCs are systematically less complete than rGCs at small cluster-centric radii in the $I$ band, while both samples are complete for distances larger than \SI{4}{arcmin}. 
The completeness curves were then parametrised using a Pritchet function \citep{McLaughlin94}. Completeness corrections were applied to the GC number density profile on a bin-by-bin basis. For each radial bin, we constructed the background-subtracted $V$-band luminosity function, corrected it for incompleteness by dividing by the corresponding completeness fraction, and fitted a Gaussian GC luminosity function (GCLF) to the bright end of the distribution ($V<$ \SI{-8.5}{mag}). The fitted Gaussian is centred at $M_{TOM} =$ \SI{-7.4}{mag}, with the dispersion left as a free parameter but constrained to be $0.8 < \sigma <$ \SI{1.5}{mag}. 
The choice for $M_{TOM}$ is a standard value for the GCLF \citep[e.g.][]{Rejkuba12}, while the choice for $\sigma$ was motivated by previous works that show that this range covers most observed galaxies \citep{Villegas10}.
The fitted GCLF is then used to estimate the total number of GCs in each radial bin, which was compared to the observed sample size to derive a multiplicative correction factor. The total GC numbers in the annuli were also corrected for geometrical incompleteness due to the detector borders and masked regions. These factors were applied to the number density profile, and the procedure was carried out independently for the full GC sample as well as for the red and blue subpopulations.}

Using Eq. \ref{eq:SN}, we calculated a localised $S_N$ by restricting the calculation to each annulus individually, thus obtaining a $S_N$ profile.
For each annulus, we obtained $N_{GC}$ by integrating the background-subtracted number density profile in that region, and $M_V$ by integrating the $r$-band surface brightness profile measured by \cite{Spavone24} and transforming to the $V$ filter\footnote{The conversion from $r$ to $V$-band magnitudes was carried out using the relation by \cite{KostovBonev18}: $M_V = M_r - 0.017 + 0.492(g-r)$, assuming a constant colour $g-r=$ \SI{0.7}{mag} motivated by the colour profiles from \cite{Spavone24}.}.
This way, what is measured is the richness of GCs with respect to the underlying stellar component inside each bin, which can be compared to typical values of $S_N$ for different galaxy types. 
These profiles will be presented in Sec. \ref{sec:SN}.

\section{Globular cluster spatial distribution and radial profiles}
\label{sec:spatial_dist}

\begin{figure}
    \centering
    \includegraphics[width=0.9\columnwidth]{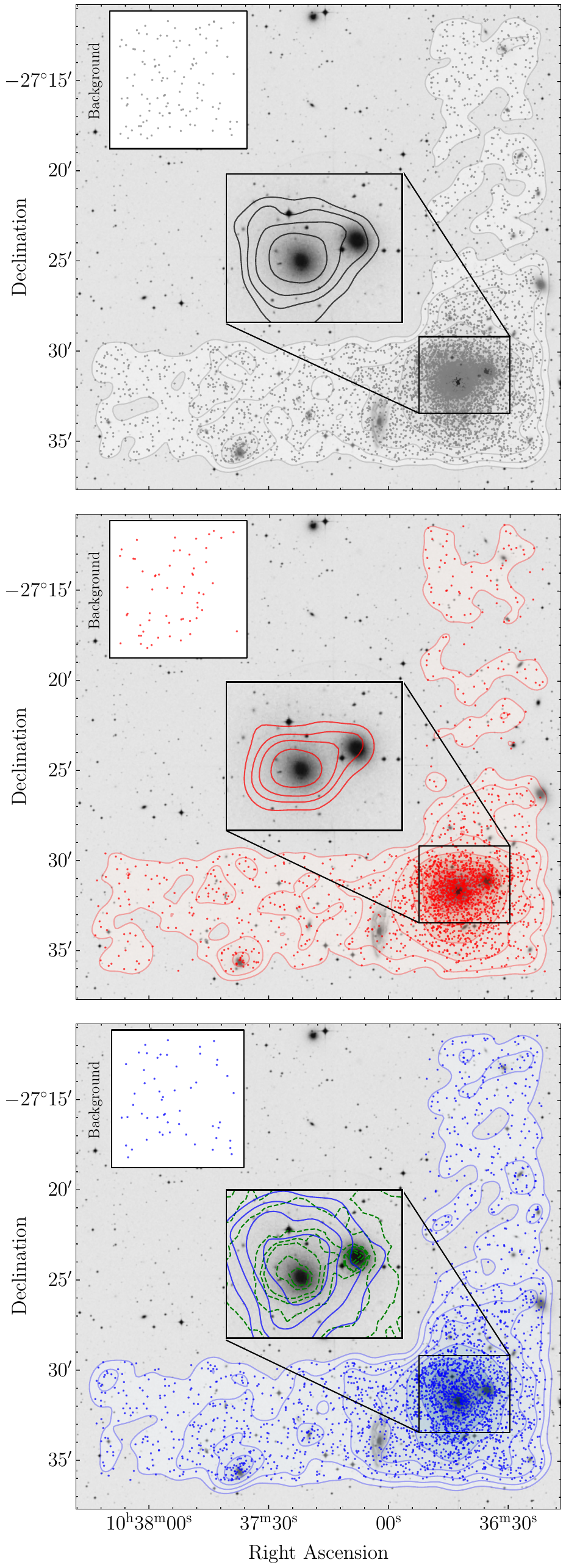}
    \caption{
    Spatial distribution of all (top panel), red (middle panel), and blue (bottom panel) GCs around Hydra\,I. The insets show the distributions around the central galaxies, NGC\,3311 and NGC\,3309. The spatial distribution of background objects for each population is shown in the top left inset in each panel. The dashed green contours on the bottom panel shows the X-ray emission from \cite{Hayakawa06}.
    }
    \label{fig:spatial_dist}
\end{figure}

Figure \ref{fig:spatial_dist} shows the spatial distribution of all (top panel), red (middle), and blue (bottom) GCs in Hydra\,I. 
The insets on the top left of each panel show the spatial distribution of the background sources selected with our GC selection methodology, which are located at a projected distance of \SI{1.2}{Mpc} from the centre of Hydra\,I.
The overall distribution shows that most GCs are located in the central regions of the cluster, around the two brightest galaxies NGC\,3311 and NGC\,3309.
Smaller but significant overdensities are also observed around other bright galaxies, in particular NGC\,3316 and NGC\,3308.

The contours in the middle and lower panels reveal that rGCs are more concentrated around the cluster centre, while bGCs have a more extended distribution.
Moreover, the distributions around the central galaxies are strikingly different, as illustrated in the zoomed-in insets: rGCs tend to be concentrated around the bright central galaxies NGC\,3311 and NGC\,3309, but bGCs are significantly offset from the central galaxy to the northeastern direction.
The distribution of bGCs coincides with a displaced X-ray halo \citep{Hayakawa06}, shown as the green dashed contours, as well as with a displaced, extended photometric component discovered by \cite{Arnaboldi2012} \citep[see also][]{Barbosa2018}.
The spatial displacements of these structures with respect to NGC\,3311 are reported in Tab. \ref{tab:dist_peak}, along with the offsets of the peak values obtained from a Gaussian kernel density estimation (KDE) model of our GCs.
We also report the position of the midpoint between NGC\,3311 and NGC\,3309, and verify that it does not coincide with any of the peaks in the GC spatial distributions.
{In fact, bGCs are even more displaced from the midpoint between the two galaxies than from NGC\,3311, as can be seen from the contours.}

\begingroup

\renewcommand{\arraystretch}{1.5}
\begin{table}[]
\small
\caption{Peak of spatial distributions with respect to NGC\,3311.}
\begin{tabular}{llll}
\hline
\hline
            & $X$ (kpc)              & $Y$ (kpc)                               \\ \hline
NGC\,3311 & $0.00$ & $0.00$ \\
NGC\,3311 - NGC\,3309 midpoint & $-11.90$ & $4.02$ \\
{bGCs} & $1.02_{-1.08}^{+1.08}$ & $3.66_{-1.18}^{+1.19}$  \\
{rGCs} & $-0.75_{-0.88}^{+0.88}$ & $1.03_{-0.58}^{+0.59}$ \\
{Extreme bGCs} & $1.38_{-3.94}^{+9.71}$ & $10.34_{-3.99}^{+6.84}$  \\
{Extreme rGCs} & $-0.42_{-1.90}^{+1.91}$ & $1.86_{-1.22}^{+1.15}$  \\ 
Extended envelope & $3.97_{-0.36}^{+0.36}$ & $6.67_{-0.45}^{+0.45}$  \\ 
X-ray secondary peak & $4.87$ & $3.49$  \\  \hline
\end{tabular}
\tablefoot{Values for GCs were obtained from a Gaussian KDE model, the extended envelope centre was taken from \cite{Barbosa2018} and adjusted to our distance modulus, and the X-ray peak was obtained visually from the contours. The confidence intervals correspond to the 16th and 84th percentiles estimated from bootstrapping with 10000 resamples. $X$ and $Y$ values increase to the north and east, respectively.}
\label{tab:dist_peak}
\end{table}
\endgroup

\begin{figure*}
    \centering
    \includegraphics[width=0.9\textwidth]{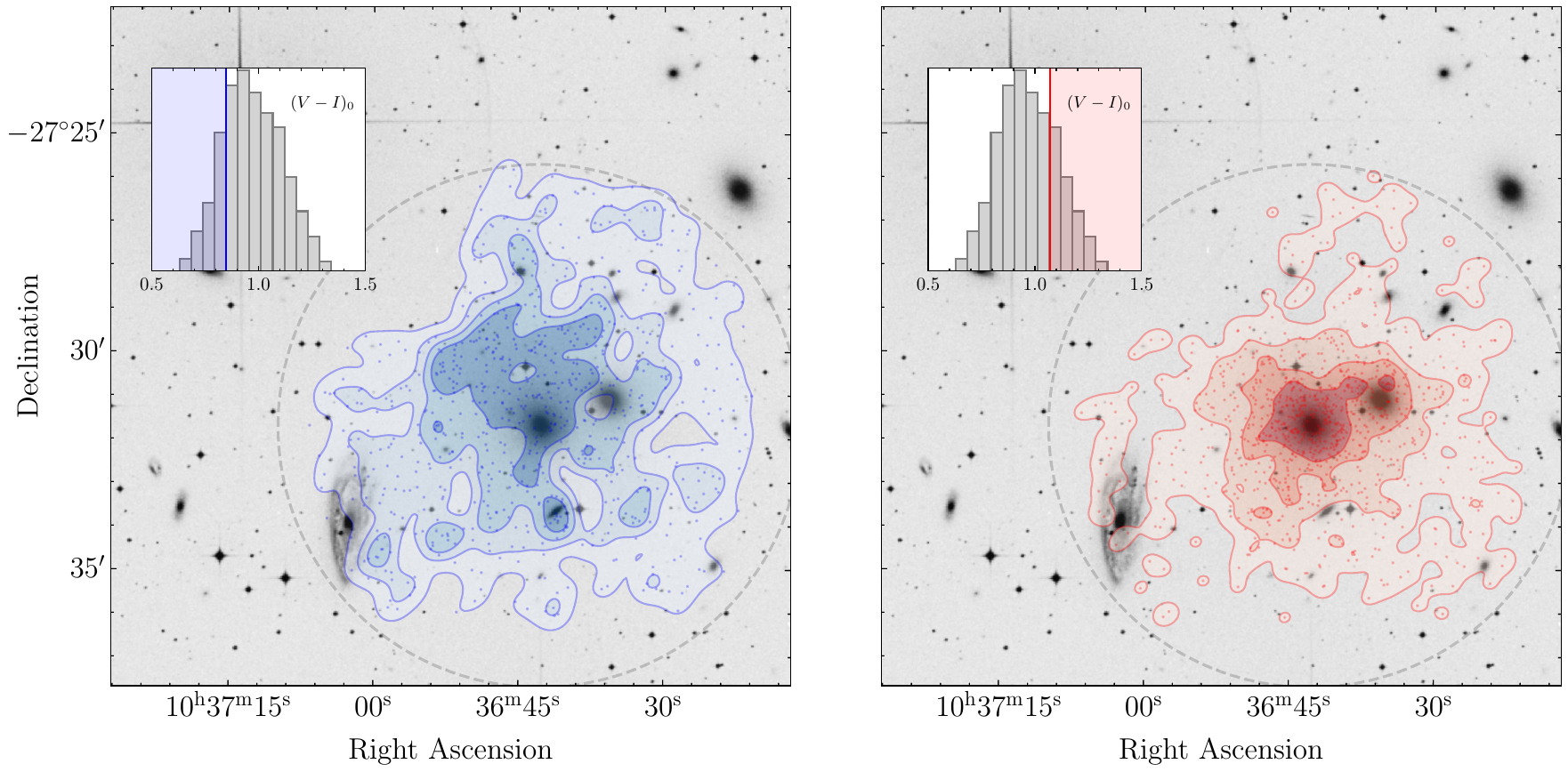}
    \caption{Central ($d \leq$ \SI{6}{arcmin}, grey dashed circle) spatial distribution of very blue (left) and very red (right) GC candidates, with colour ranges indicated in the insets.}
    \label{fig:extremes_dist}
\end{figure*}

It is also interesting to study the central ($d<$ \SI{6}{arcmin}) distribution of extreme populations of bGCs and rGCs, that is, the very blue ($(V-I)_0 \lesssim 0.8$) and very red ($(V-I)_0 \gtrsim 1.1$) populations.
Figure \ref{fig:extremes_dist} and Tab. \ref{tab:dist_peak} show that the extreme bGCs are even more displaced from the central galaxy than the general blue population, while the extreme rGCs seem to be mostly associated with NGC\,3311.
Moreover, they seem to be slightly displaced to the southwest of the galaxy, opposite to the bGCs {(see contours in Fig. \ref{fig:extremes_dist})}.
We verify that this result is present even if we consider only GCs with very small photometric errors ($<$ \SI{0.05}{mag}).
The opposite displacement of the extreme populations with respect to NGC\,3311 may be additional evidence for the sloshing of NGC\,3311 in the cluster's dark matter halo suggested by \cite{Barbosa2018}.
The less bound bGCs are more easily displaced as NGC\,3311 moves through the dark matter halo, while the more bound rGCs are only slightly displaced in this process.
This process will be studied in detail in conjunction with idealised N-body simulations in a future work.

\subsection{GC stellar populations}
\begin{figure}
    \centering
    \includegraphics[width=\columnwidth]{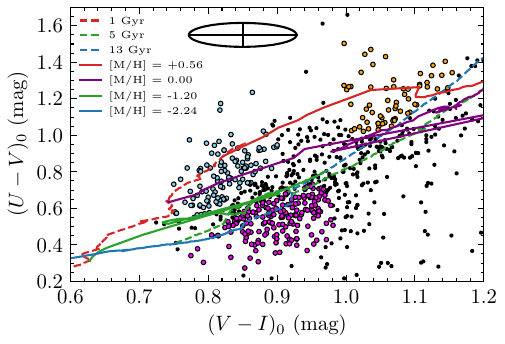}
    \caption{$UVI$ colour-magnitude diagram of GC candidates with photometric error $\leq$ \SI{0.07}{mag} in each band (dots). The lines correspond to stellar tracks of constant age (dashed) and constant metallicity (solid) derived from the PARSEC models \citep{Bressan2012}. The black ellipse represents the mean errors in colour.
    We select three subpopulations of GCs based on the PARSEC tracks: young, metal-rich (light blue), old, metal-rich (orange), and old, metal-poor (magenta).
    }
    \label{fig:UVI_cmd}
\end{figure}

It is important to understand the contributions of GCs with different stellar populations to the overall GC distribution, and how their spatial distributions may differ from one another.
In the central regions, the $U$-band data allows us to separate our GC sample in young and old, metal-rich and poor subsamples.
Figure \ref{fig:UVI_cmd} shows the extinction-corrected $(U-V)_0$ and $(V-I)_0$ colours of our GC candidates, with the additional restriction of having a low photometric error ($<$ \SI{0.07}{mag}) in all bands.
Overplotted are tracks of constant metallicity (solid lines) and constant age (dashed lines) resulting from the PARSEC stellar population models \citep{Bressan2012} that serve as guides for the age and metallicity locations in this plane.
We isolate three subpopulations of GC candidates: young, metal-rich (in light blue), old, metal-poor (in magenta), and old, metal-rich (in orange).
{They represent 15\%, 24\%, and 8\% of the low-error $UVI$ sample, respectively.}
By inspecting the distribution of background objects in this colour-colour space, we verified that the contamination is insignificant, with only two contaminants present in the young, metal-rich region.

\begin{figure*}
    \centering
    \includegraphics[width=\textwidth]{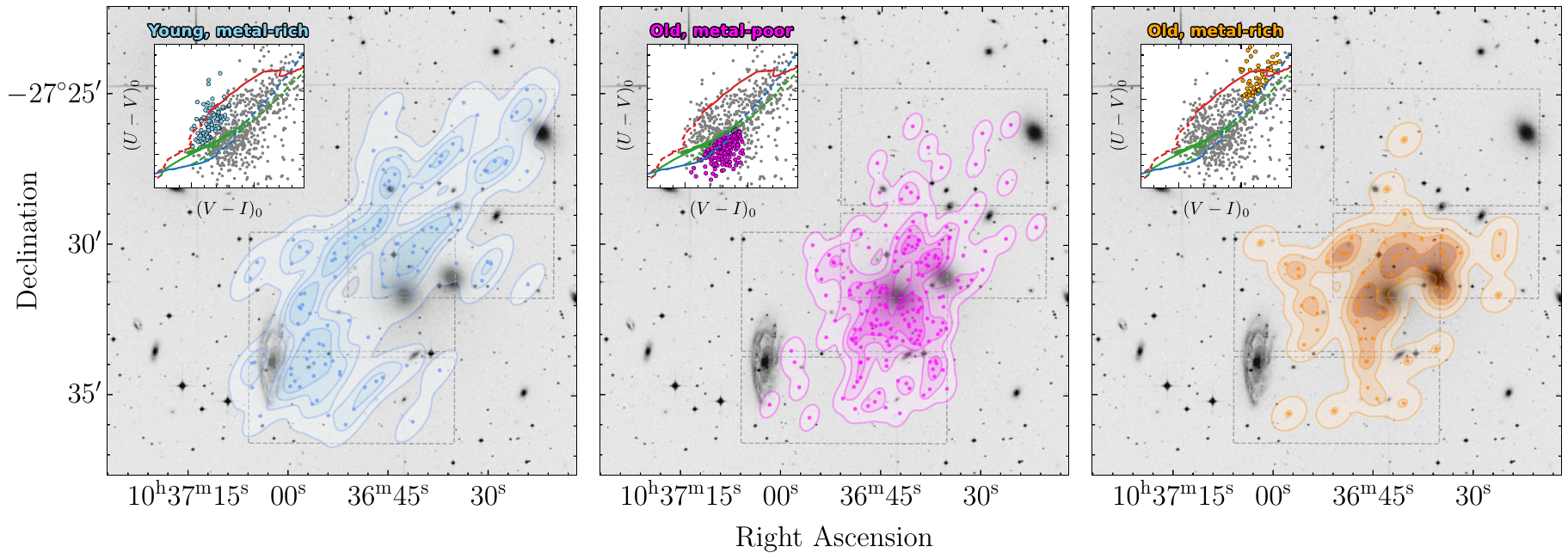}
    \caption{Spatial distribution of the three subpopulations from Fig. \ref{fig:UVI_cmd}. From left to right: young metal-rich GCs, old metal-poor GCs, and old metal-rich GCs. The $U$ field coverage is indicated as dashed grey lines.}
    \label{fig:UVI_dist}
\end{figure*}

These subpopulations display very different spatial distributions, as shown in Fig. \ref{fig:UVI_dist}.
The young, metal-rich population dominates the displaced population of bGCs, and their distribution is extended to the south-east of NGC\,3311, suggesting a possible association with the ram-pressure tails of NGC\,3312 revealed by $\textrm{H}\scriptstyle\mathrm{I}$ data \citep{Hess22}.
The old metal-poor population shows different substructures.
It is slightly offset from the central galaxies, with one overdensity to the north and another to the south of NGC\,3311.
The southern overdensity is particularly interesting, as it coincides spatially with the tidal tails emerging from HCC\,007, a low-mass S0 galaxy being disrupted by the BCG \citep{Arnaboldi2012}.
Old metal-poor GCs are expected to be found in S0 galaxies, demonstrating the power of GCs in tracing faint stellar substructures.
Finally, the old, metal-rich GCs are not as displaced from the centre and seem to connect the two bright central galaxies.
Such old, metal-rich GCs are expected to form in situ in massive galaxies, and thus are likely associated with NGC\,3311 and NGC\,3309 \citep{BrodieStrader06}.
In what follows, the analysis is done on the $VI$ GC candidates without a distinction in stellar populations, leveraging the wider spatial extension of this dataset.

\subsection{GC number density profiles and comparison with BCG+ICL surface brightness}
\begin{figure*}
    \centering
    \includegraphics[width=\textwidth]{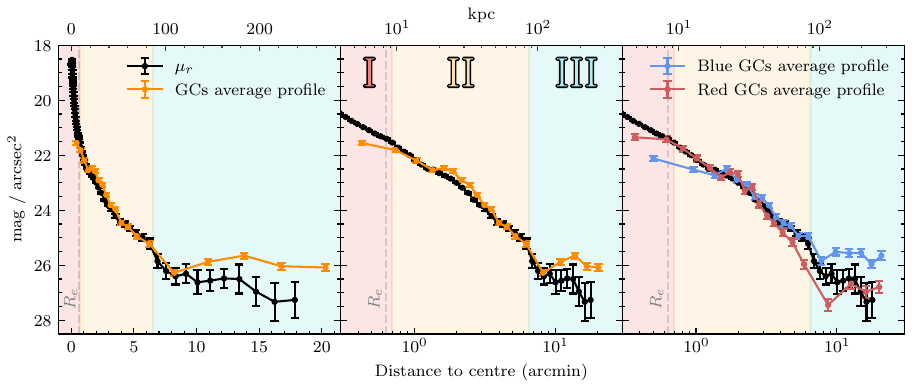}
    \caption{Number density profile of GCs and the surface brightness profile of NGC\,3311 (black points). The left and central panels show the number density profile considering all GC candidates (orange points) with a linear and logarithmic distance scale, respectively. The three different regimes of the GC number density are indicated with filled colours, and are more clearly distinguished in the middle panel. The right panel shows the number density profiles of red and blue GCs separately, and showcases how the red GCs follow the starlight while blue GCs follow a faint, diffuse component. All the number density profiles were arbitrarily shifted to best match the surface brightness profile in the central regime. The grey dashed line indicates the effective radius of NGC\,3311.}
    \label{fig:num_dens}
\end{figure*}

The average number density profile of all $VI$-selected GCs is compared to the surface brightness profile of NGC\,3311 \citep{Spavone24} in the left and middle panels of Fig. \ref{fig:num_dens}. 
The GC profile is arbitrarily shifted to match the surface brightness and we indicate three regions of interest that are analysed separately.
Region I covers {small clustercentric distances ($d\lesssim1.1\,R_e$)} and shows a flatter distribution of GCs with respect to the stellar profile. 
This mismatch is observed both in the total population and in rGCs and bGCs separately (right panel). 
{At these small distances, our GC sample still suffers from incompleteness effects due to the high surface brightness of NGC\,3311, which makes a reliable extraction of point sources challenging.}
Region II shows the region where the GC number density closely follows the stellar spheroid and inner halo profile of NGC\,3311 {extending from $0.7 \leq d \leq 6.5$ arcmin (i.e. $1.1 \,R_e \leq d \leq 10.3\,R_e$).}

Interestingly, {region III ($d \geq$ \SI{6.5}{arcmin})} shows a significant difference between the GC number density and the surface brightness profile. 
While the surface brightness continues to decrease, the GC profile flattens. 
This indicates a change in the richness of GCs over the observed stellar component and marks a transition between the stellar halo and ICL dominated regions. 
This is in agreement with the photometric decomposition presented in \cite{Spavone24} (double S\'ersic + exponential) that showed an extended exponential profile starting to dominate at $d \sim 6$ arcmin.

The right panel of Fig. \ref{fig:num_dens} shows the number density of rGCs and bGCs separately, and reveals an important difference between the two populations. 
The rGCs closely follow the galaxy light from NGC\,3311 out to large clustercentric distances, while bGCs trace a more extended component. 
It is clear that the bGCs are responsible for the shallower decrease of the total GCs number density profile observed in region III.
There is also a clear difference between rGCs and bGCs in region I. 
The distribution of rGCs follows the galaxy light very closely down to a few hundred parsecs, with the aforementioned incompleteness effects only taking place in very small distances from the galaxy centre.
However, the bGCs profile remains below the surface brightness up to $\sim$ \SI{1.5}{arcmin}. 
This discrepancy {reflects} the fact that bGCs originate mainly from low mass, metal-poor systems that were tidally disrupted and deposited their stars and clusters at larger distances, while rGCs tend to be metal-rich and be formed in more massive galaxies, including NGC\,3311 itself \citep{Hilker99, Ramos2018}.
The ex situ rGCs can reach the centre more easily in merger events \citep{Amorisco17}.
Additionally, GC destruction processes acting in the very centre of massive galaxies may also play a role in the observed flattened number density profile \citep{Goudfrooij07}, with N-body simulations showing that less compact clusters get disrupted after 2-3 passages through the galactic centre \citep{Miocchi2006}.

It may be surprising how well bGCs trace the halo of NGC\,3311, with both profiles closely following each other in region II.
In other clusters, the bGCs profile is usually shallower than the surface brightness at all radii \citep[e.g.][]{Cantiello2018}.
Since the tagging of GCs as 'red' and 'blue' was done in a statistical way, some GCs may be misidentified as blue, especially around the large overlap at $(V-I)_0 \sim $ \SI{0.9}{mag} (Fig. \ref{fig:gmm}).
By measuring the number density profile of only very blue GCs ($(V-I)_0 \lesssim$ \SI{0.8}{mag}) we indeed verify that this profile follows a single power law, shallower than the surface brightness profile.

\subsection{Specific frequency variations as function of cluster radius}
\label{sec:SN}

\begin{figure}
   \centering
   \includegraphics[width=\columnwidth]{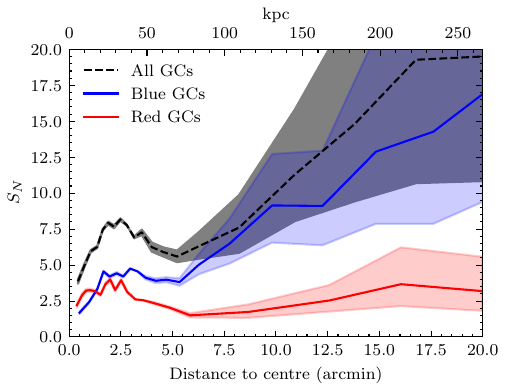}
   \caption{Local specific frequency profile of our GC candidates. Red GCs maintain a lower specific frequency, typical of bright galaxies, while blue GCs reach values typical of dwarf galaxies.}
    \label{fig:SN}
\end{figure}

The localised specific frequency profile obtained from the GC number density and stellar surface brightness profiles, $S_N$, is shown in Fig. \ref{fig:SN}.
{The specific frequency of rGCs is nearly constant at $\sim 2.5$ for all radii}, reflecting the fact that the rGC number density follows the surface brightness profile.
The bGCs have a comparable $S_N$ to that of rGCs up to $\sim$ \SI{5}{arcmin}, after which a {sharp increase is observed to $S_N \sim 15$.}
This behaviour indicates a significant change in the mixture of bGCs and stars as we move away from the centre of Hydra\,I, as also indicated by their number density profile.
Similar trends were reported for NGC\,1399 in the Fornax cluster \citep{Cantiello2018, Cantiello2020}.
The high $S_N$ measured at large distances is typical of dwarf galaxies \citep{Georgiev2010}, and would therefore be expected if the stellar material at these distances was deposited there following the disruption of these less massive systems. This is in line with studies on resolved stellar population in the Virgo cluster by \cite{Williams2007} that indicate that the ICL was built primarily through the disruption of dwarf galaxies, as 70\% of the stars in their intracluster HST field have metallicity [M/H] $\sim -1.3$. 

\begin{figure}
    \centering
    \includegraphics[width=\columnwidth]{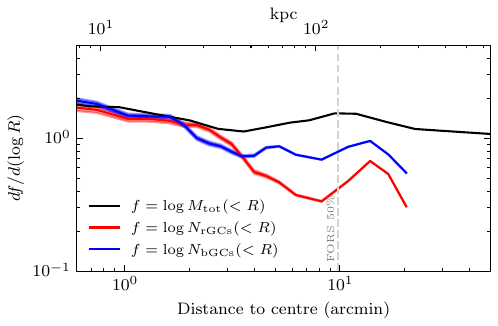}
    \caption{Logarithmic slope of the total cumulative mass profile (black) modelled by \cite{Hayakawa06} from X-ray observations compared to that derived from the cumulative number of GCs (red and blue). The dashed line shows the distance where the coverage of our data goes down to 50\% of the X-ray coverage at the same distance.}
    \label{fig:cum_N_GC}
\end{figure}

Both the number density and $S_N$ profiles suggest that bGCs are connected to the ICL, and therefore to the galaxy cluster potential.
In addition, the number of GCs in a galaxy system is a robust tracer of the underlying total mass \citep[and references therein]{BurkertForbes20}.
In Fig. \ref{fig:cum_N_GC} we show the slopes of three radial profiles: the cumulative dynamical mass modelled from X-ray observations in \cite{Hayakawa06} and the cumulative number of rGCs and bGCs. 
At small radii, both rGCs and bGCs have comparable slopes to the total mass, but for larger distances both populations show a shallower profile, although we verify that bGCs trace the total mass more closely than rGCs.
This supports the prevalence of an intracluster population dominated by bGCs that follows the total gravitational potential \citep{Forbes2012}.
As a caveat, we should bear in mind that the mass model is derived from the hot gas temperature assuming hydrostatic equilibrium, which is not accurate in the case of shocks caused by cluster mergers that lead to overestimated mass measurements and artificial slopes \citep{Nelson12}.
In fact, \cite{Hayakawa06} does detect a slight temperature excess \SI{7}{arcmin} southeast of NGC\,3311 that could hint towards a past subcluster merger.
Finally, the partial spatial coverage of our data compared to the X-rays for radii $\gtrsim$ \SI{10}{arcmin} may also affect the shape of the GC slope profiles and how they compare to the total mass profile, where the coverage from FORS is lower than 50\% of the X-ray coverage.

\section{Constraining the past galaxy luminosity function of Hydra\,I}
\label{sec:LF_evo}

The two-phase formation scenario for galaxies \citep{Oser10} predicts that a significant fraction of stellar halos in central cluster galaxies can originate from accreted stars during galaxy interaction and merger events \citep{Pulsoni21}.
For example, the numerical models introduced in \cite{Amorisco17} reveal that the stars coming from more massive mergers are mostly deposited at smaller radii, while those coming from less massive ones cannot easily sink into the potential well and thus deposit stars in the outskirts.
Similarly, the dominant formation channel for the ICL is through disruption and merger events, with a small contribution of in situ star formation \citep{Rudick09}.
A similar formation channel is expected for intracluster GCs, and therefore, in terms of colour and specific frequencies, the intracluster GC population should be representative of the former galaxy populations of the cluster that were destroyed during these galaxy interactions \citep{Cote98, Ahvazi24}.

The scenario where metal-poor GCs come mainly from disrupted dwarf galaxies can be tested against theoretical predictions of the tidal stripping of these systems.
In Appendix \ref{appendix:tidal_disruption}, we present a simple argument based on the tidal radius of dwarf galaxies to derive the clustercentric distance where these systems start to be significantly affected by tidal disruption.
We argue that less massive systems start losing most of their stars at large distances ($\sim$ \SI{200}{kpc} for $m=$ \SI{e7}{M_\odot}), while more massive ones sink deeper in the cluster before losing their stars to the intracluster space ($\sim$ \SI{50}{kpc} for $m=$ \SI{e8}{M_\odot}).

\cite{Amorisco17} shows that dynamical friction can efficiently radialise only the orbits of stars coming from the more massive mergers.
In other words, the stars from less massive satellites that have a high circularity at infall tend to remain at large radii for longer.
Thus, the dwarf galaxies at the distances shown in Fig. \ref{fig:tidal_rad} can contribute significantly to the intracluster population of metal-poor GCs, explaining the high specific frequency measured in Hydra\,I and other clusters. 

In this framework, we can try to use the measured properties of GCs to constrain the galaxy luminosity function of Hydra\,I as it was before the build-up of the currently observed ICL.
In particular, we use the GC specific frequency profile of Fig. \ref{fig:SN} to constrain the faint end of the past luminosity function, as this parameter is sensitive to disrupted dwarf galaxies that have typically rich GC systems. 
The bright end will be constrained according to the GC colour distribution, since the massive galaxies that merged to form the cluster centre have well defined and distinct GC colour distributions.

\subsection{Constraining the faint end}
\label{sec:faint-end-constr}
The GC specific frequency of a galaxy is a quantitative measure of the number of GCs in comparison to the galaxies' stellar light, and has a known dependence on galaxy luminosity \citep{Harris91, Georgiev2010}.
High specific frequencies are found both in very bright ($M_V \sim$ \SI{-22}{mag}) and dwarf ($M_V \gtrsim$ \SI{-18}{mag}) galaxies, while the lowest values are common in average luminosity galaxies ($M_V \sim$ \SI{-20}{mag}). 
The high specific frequency of the ICL calculated at large distances (Fig. \ref{fig:SN}) could be therefore representative of the dwarf galaxies that were tidally destroyed during the assembly history of the cluster.
We can translate our $S_N$ profile into typical galaxy luminosities using the following scaling relation introduced by \cite{Georgiev2010},
\begin{equation}
\label{eq:SN_MV}
    S_N = \frac{\eta 10^6}{m_{TO}}\left(\kappa_1^{0.6}10^{0.16M_V+0.24M_{V,\odot}} + \kappa_2^{2}10^{-0.4M_V+0.8M_{V,\odot}}\right)\, ,
\end{equation}

where $\eta=5.5\times10^{-5}$ is the GC formation efficiency, $m_{TO}=1.69\times10^5M_\odot$ is the turn-over mass, $\kappa_1=1\times10^9$ and $\kappa_2=10^{-4.42}$ are normalisation parameters \citep[see Eqs. 11 and 12 in][]{Georgiev2010} and $M_{V, \odot}=$ \SI{4.82}{mag} is the solar $V$-band magnitude. 
This equation reproduces the characteristic U-shape relation between $S_N$ and $M_V$ (see Fig. 6 in \citealt{Georgiev2010}), with the faint and bright ends of this relation being dominated by the first and second terms, respectively.
This way, for a given radial bin, we can obtain the typical galaxy magnitude $M_V$ that corresponds to the specific frequency that we measure in that bin.
We stress that we only use the faint regime of Eq. \ref{eq:SN_MV} ($M_V>-20$ mag) to convert $S_N$ into $M_V$ since, for large clustercentric distances, the high $S_N$ likely comes from the disruption of dwarf galaxies.
Dividing the amount of light inside each bin, obtained by integrating the surface brightness profile, by the luminosity corresponding to $M_V$, we obtain the average number of dwarf galaxies of that magnitude that had to be disrupted in order to build the observed BCG+ICL luminosity.
Lastly, we add the number of destroyed dwarfs to the present-day luminosity function to obtain a prediction of the luminosity function in place at an early assembly stage of the cluster. 
The top panel of Fig. \ref{fig:LF_constraint} shows the present-day luminosity function for dwarf galaxies in Hydra\,I based on the catalogue from \cite{LaMarca2022}. 
It is well described by a Schechter function with {slope $\alpha=-1.41_{-0.05}^{+0.08}$}, comparable to that of the Virgo cluster \citep{Morgan25}.
With the addition of the destroyed dwarf galaxies, the luminosity function is best described by a Schechter function with a {steeper slope of $\alpha=-1.97_{-0.22}^{+0.24}$.}
We repeated the same exercise considering the $S_N$ of bGCs only, and found a similar slope of $\alpha=-1.81_{-0.16}^{+0.16}$.
Following Eq. \ref{eq:SN_MV}, the lower bGC $S_N$ values compared to the total $S_N$ profile yield progenitors with a higher contribution from bright galaxies, although in both cases the faint-end slopes are consistent.
This is because the faint end is sensitive to {the annuli with high ($\gtrsim 5$) $S_N$}, typical of low mass galaxies, which are obtained for both the total and the blue GC samples.
The best fit parameters were obtained using a simple MCMC model fitting approach, assuming Gaussian likelihoods and flat priors, and are shown in Tab. \ref{tab:faint_end_fit}.

{Note that our method would produce a delta function in the past LF in the case of a perfectly constant $S_N$ profile.
We discuss this case in more detail in Appendix \ref{appendix:faint_end_constant_sn}.
Additionally, }our approach makes two important simplifications: that every value of $S_N$ maps to a single progenitor galaxy with $M_V$, and that every galaxy contains GCs (i.e. the GC occupation fraction is $f_\textrm{GCs}=1$).
In reality, there is significant scatter in the $S_N$-$M_V$ relation (Fig. 6 in \citealt{Georgiev2010}), and there is a known dependence of $f_\textrm{GCs}$ with galaxy magnitude.
We explore a refined method of constraining the faint end of the LF taking these effects into account in Appendix \ref{appendix:faint_end_refined}, but include our findings in the top panel of Fig. \ref{fig:LF_constraint} as golden points and in Tab. \ref{tab:faint_end_fit}.
Now, the constrained {faint-end slope is $\alpha=-1.81_{-0.13}^{+0.10}$}, less steep than the valued obtained with the simplified method but still significantly steeper than the present-day value and with a tighter fit.

\begingroup

\renewcommand{\arraystretch}{1.5} 
\begin{table}[]
\caption{Best-fit Schechter parameters of the derived luminosity functions.}
\begin{tabular}{llll}
\hline
\hline
            & $\phi^\star$              & $M^\star$ (mag)               & $\alpha$               \\ \hline
{Present-day} & $6.15_{-4.36}^{+8.04}$ & $-20.16_{-2.77}^{+1.34}$ & $-1.41_{-0.05}^{+0.08}$ \\
{Past, all GCs}        & $67.90_{-54.31}^{+125.81}$ & $-17.47_{-1.04}^{+0.59}$ & $-1.97_{-0.22}^{+0.24}$ \\
{Past, bGCs}        & $98.65_{-70.72}^{+110.42}$ & $-17.64_{-0.95}^{+0.52}$ & $-1.81_{-0.16}^{+0.16}$ \\ 
{Past, Gauss}     & $169.20_{-96.19}^{+85.38}$ & $-17.24_{-0.54}^{+0.26}$ & $-1.81_{-0.13}^{+0.10}$ \\ \hline
\end{tabular}
\tablefoot{Reported values are the medians of the respective posterior distribution, while the uncertainties correspond to the 16th and 84th percentiles.
}
\label{tab:faint_end_fit}
\end{table}
\endgroup

\subsection{Constraining the bright end}
The bright end of the past luminosity function is studied from a different angle.
While the high specific frequencies measured at large distances are sensitive to the past dwarf galaxy population, the colour distribution of GCs can be used to trace the bright galaxies that merged to build the currently observed central galaxies.
Using the ACS Virgo Cluster Survey \citep{Cote04}, \cite{Peng2006} have demonstrated that the colour distribution of GCs changes significantly as a function of the host galaxy magnitude.
In particular, bright massive galaxies show a double peak in colour, while dwarf galaxies exhibit a single blue peak with a red tail.
These colour distributions are reproduced in Appendix \ref{appendix:templates}.

\begin{figure}
    \centering
    \includegraphics[width=\columnwidth]{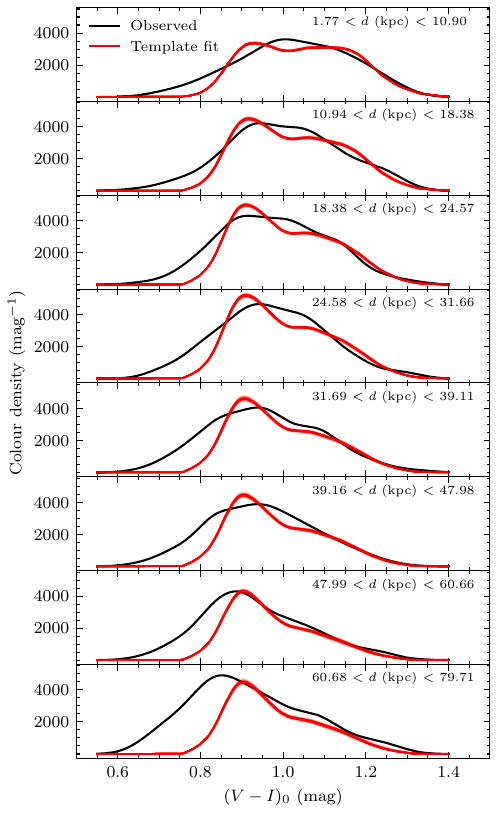}
    \caption{Colour distribution of GCs in radial bins. Black lines show the observed distribution in Hydra\,I, while the red dashed lines show the best fit obtained from the ACS templates.}
    \label{fig:template_fit}
\end{figure}

We attempt to use the Virgo GC colour profiles as a library of templates to find the best fit to the GC colour distribution in Hydra\,I, under the strong assumption that Hydra\,I has a similar assembly history as Virgo.
This is done by fitting the colour distribution in radial bins independently, up to a distance of \SI{6}{arcmin}, which is the region where the light is dominated by the BCG.
Figure \ref{fig:template_fit} shows the observed GC colour distributions in each bin.
The total counts were corrected for photometric incompleteness and accounted for the missing area that is not covered by the FORS fields.
Additionally, the colour distribution of the background field was scaled to the area of the annulus and subtracted from the GC colour distribution.
The best fit obtained from a combination of the ACS colour templates are also displayed.
Notice how the observed distribution shifts to blue-dominated GCs as we move away from the centre, suggesting an increasingly large contribution from less massive galaxies to the GC population at larger distances.

The ACS templates fit reasonably well the colour distribution in Hydra\,I for the red part and the central rings ($d\leq$ \SI{20}{kpc}), however it shows some discrepancies at larger radii. 
In particular, the templates cannot reproduce the population with very blue colours ($(V-I)_0 \lesssim$ \SI{0.8}{mag}), indicating that such GC systems are not present in the Virgo galaxies studied in the ACS survey.
This discrepancy may be related to the different dynamical state of the two clusters.

\begin{figure}
    \centering
    \includegraphics[width=\columnwidth]{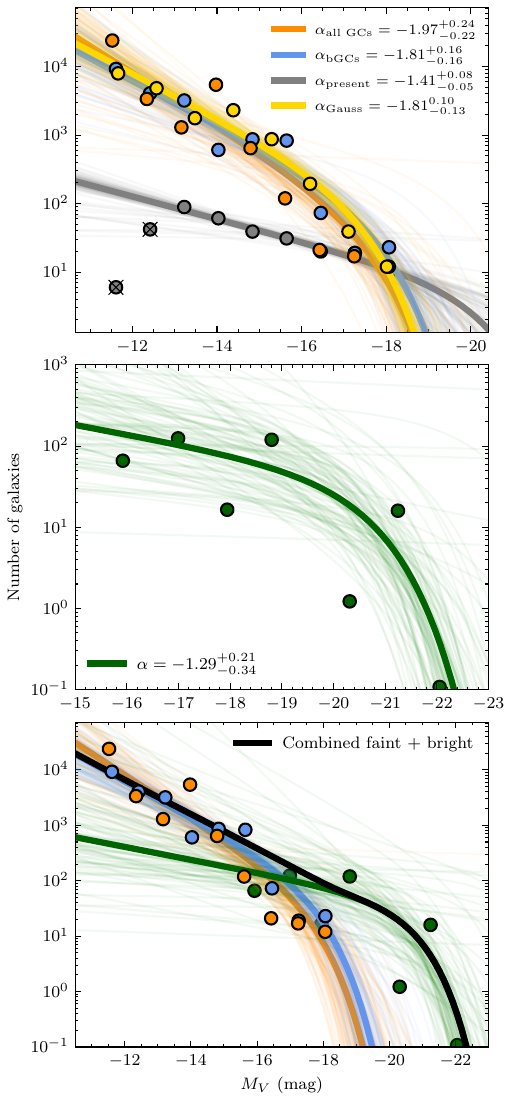}
    \caption{Constraints to the past luminosity function of Hydra\,I. Top: The faint end of the luminosity function. As grey points, we plot the data from the dwarf galaxy census by \cite{LaMarca2022}, while in orange and blue we plot the predicted number of destroyed dwarf galaxies added to the present-day LF from all GCs and bGCs only, respectively. 
    The golden points show the constraint taking into account the GC occupation factor and the scatter in magnitude from the $S_N$-$M_V$ relation.
    Thick lines show the Schechter fit with maximum likelihood to our data, while thin lines represent random draws from our MCMC posteriors. Crossed data points were not considered for their respective Schechter fit as they suffer from incompleteness. 
    Middle: Bright-end luminosity function prediction from fitting GC colour templates to the observed colour distributions from all radial bins.
    Bottom: combined constrain from both methods, considering the blue curve for the faint-end.}
    \label{fig:LF_constraint}
\end{figure}

Each bin is fitted with a (linear) combination of colour templates, which allows us to extract the fractional contribution of each template in each bin.
We can thus obtain the total number of galaxies that contributed to the observed GC population by stacking all radial bins.
The middle panel of Fig.  \ref{fig:LF_constraint} shows the total number of galaxies from each colour template as a function of their characteristic absolute magnitude, with the best fit of a Schechter function.
{We find that the expected downward trend is recovered with a Schechter slope of $\alpha=-1.29_{-0.34}^{+0.21}$.}
Finally, the two constraining methods are combined in the bottom panel of Fig. \ref{fig:LF_constraint}, shown as the black curve.

Another consistency check for the obtained bright-end is to compare the total luminosity of these galaxies to that observed in the present day inside \SI{6}{arcmin} radius.
The present-day total luminosity is obtained by integrating the surface brightness profile, and adding the luminosities of the masked galaxies inside the aperture (namely: NGC\,3307, NGC\,3309, NGC\,3312, HCC\,006, HCC\,007, HCC\,009, HCC\,011, HCC\,012, and HCC\,013).
{This results in a total present-day magnitude of $M_r = $ \SI{-24.13}{mag}.
The total magnitude estimated from the template fitting is $M_r = $ \SI{-25.37}{mag}.
This corresponds to a factor of $\sim 3.1$ more flux predicted by the colour distribution of GCs.}
The excess luminosity may partly be explained by the mismatches with the Virgo templates, and by stars being distributed at distances larger than \SI{6}{arcmin} during the hierarchical assembly of the cluster.

\section{Discussion}
\label{sec:discussion}

The GC population in Hydra\,I reveals the presence of substructures that suggest that the cluster is in an active assembly stage.
The different spatial distributions and specific frequencies of the red and blue GC populations also provide constraints for the hierarchical assembly of Hydra\,I.
Similar trends have been reported for other groups and clusters, which are now discussed.

The displacement of GCs in galaxy clusters with respect to the BCG is not uncommon and is expected for unrelaxed systems. 
Recently, \cite{Kluge25} analysed the GC population in the Perseus cluster using data from the Euclid Early Release Observations and showed that ICGCs closely trace the ICL, reporting a displacement of \SI{60}{kpc} between the distribution of ICGCs and the BCG.
In the Dorado group, \cite{Urbano25} found that while rGCs are centred on the group members, a significant offset is observed for bGCs, similar to what we observe in Hydra\,I. 
Similar results were reported for NGC\,1399 in the Fornax cluster \citep{Cantiello2020}.

The different number density profiles of rGCs and bGCs clearly show that the two populations are tracing distinct structures at large radii.
In the Virgo cluster, \cite{Durrell14} found similar trends for M87 and M49 as we see in Hydra\,I: rGCs closely follow the surface brightness profile and are located around massive galaxies, while bGCs have a much more extended spatial distribution \citep[see also][]{Lee10}.
Similar trends have been reported for other Virgo galaxies (M49, \citealt{Lee98}; M60, \citealt{Forbes04}; M89, \citealt{Tamura06}), for NGC\,1399 \citep{Schuberth10, Cantiello2020} and NGC\,3115 \citep{Cantiello2018}, as well as for ICGCs in the Coma cluster \citep{Madrid18}.
These trends also seem to hold in the less dense environments of dynamically young galaxy groups \citep[NGC\,5018,][]{Spavone18, Lonare25}
Additionally, \cite{Longobardi15a} have confirmed that PNe associated with the halo of M87 follow a S\'ersic radial profile, matching the stellar component, while ICPNe follow a shallow power law profile.

The rising specific frequency profile of bGCs (Fig. \ref{fig:SN}) indicates that the observed BCG+ICL light is typical of massive (dwarf) galaxies at smaller (larger) clustercentric distances.
A similar trend was found by \cite{Kluge25} for the ICGCs in the Perseus cluster.
Similarly, \cite{Cantiello2018} studied the GC population of NGC\,1399 in the Fornax cluster and of the isolated lenticular galaxy NGC\,3115, and found that the cumulative specific frequency $S_N(<R)$ in NGC\,1399 shows the same trends as reported in Fig. \ref{fig:SN}, with rGCs remaining constant at low values, and a significant increase for bGCs at large radii.
Interestingly, this trend is not as strong in NGC\,3115, where both populations maintain a low $S_N$ for all radii, with only a mild increase for bGCs.
This may be connected to the different environments where the two galaxies are located.
NGC\,1399 sits in the centre of the Fornax cluster, where the tidal fields are strong and able to disrupt less massive systems that contribute to the metal-poor GC population.
On the other hand, NGC\,3115 is found in an isolated environment where an intracluster component is not present.
These trends are also in agreement with predictions from numerical simulations that model the formation of the intracluster component of GCs \citep{Bekki03}.

The young, metal-rich GCs in Fig. \ref{fig:UVI_dist} seem to dominate the displaced population of bGCs and show two interesting overdensities.
One to the southwest of NGC\,3311 that could be associated with the ram-pressure tails of NGC\,3312, and another to the north of NGC\,3311.
The latter is cospatial with a group of dwarf galaxies that are kinematically detached from the cluster centre \citep{Ventimiglia11}.
Although these GCs are expected to be more common in massive systems, they can also be formed in starburst tidal dwarf galaxies \citep{Fensch19}.
Old, metal-poor GCs, on the other hand, seem to coincide with a stellar stream connecting the dwarf galaxy HCC\,007 and NGC\,3311.
Still, radial velocity measurements are needed to confirm the association of both subpopulations of GCs with their respective cospatial stellar substructures.

The method for studying evolution of the faint-end slope of the galaxy luminosity function presented in Sec. \ref{sec:LF_evo} offers an interesting avenue to understand the assembly of galaxy clusters.
By postulating that the bulk of the population of bGCs comes from dwarf galaxies, {we derive a faint-end slope of $\alpha=-1.81_{-0.16}^{+0.16}$} for Hydra\,I prior to the disruption of galaxies that contributed their GCs to the ICL.
Using HST data, \cite{McLeod21} have reported that such value is consistent with observed clusters at $z\sim3$, while JWST data suggests a higher redshift of $\sim 5$ \citep{Navarro-Carrera24}, although lower redshift slope values remain inside our error bars. 
We should bear in mind that not all ICL is build from dwarf galaxies, which -- due to their high $S_N$ -- contribute a higher fraction of GCs relative to stars \citep{Ahvazi24}.

As demonstrated in Fig. \ref{fig:UVI_cmd}, the blue population in $VI$ is a mixture of young and old GCs. 
One may raise the question of how the young GC population impacts our faint-end slope constraint, since it is based on the assumption that ICGCs come from dwarf galaxies where an old population is expected.
Since it is not possible to separate young from old in the $VI$ sample, we perform a simple test by repeating the analysis with a sharp cut in $V-I$ colour, including only GCs with $(V-I)_0>0.8$ \SI{}{mag}.
{This corresponds to $\sim 84\%$ of our GC sample.}
The result of this test is shown in Appendix \ref{appendix:faint_end_test}, {and we find that the faint-end slope is not well constrained in this condition. 
Although the obtained median slopes are consistent, we probe a limited region of the galaxy LF resulting in large error bars.} The very blue GCs, young or old, are therefore essential to probe the very faint galaxies that contributed to the formation of intracluster populations. 

The constrains to the bright end of the galaxy luminosity function from the colour distribution of GCs is more subtle and presents further caveats. 
This method may only be viable when comparing clusters with a similar dynamical state, which may not be the case for Virgo and Hydra\,I, as suggested by the excess of very blue GCs in Hydra\,I\footnote{{\cite{Peng2006} also provides the median blue GC colour for different galaxy magnitudes (their Fig. 3). They report a mild evolution with magnitude, where low-mass galaxies show bluer peaks. However, even extrapolating this relation to $M_B\sim-13$ mag galaxies, we still do not predict blue colours as observed in Hydra\,I. Thus a significant fraction of the very blue GC population is consistent with intermediate-age stellar populations rather than exclusively old, metal-poor clusters, indicating a younger dynamical state of the cluster.}}.
Furthermore, the ACS Virgo Survey presents more precise photometry with smaller errors than our FORS images, which makes a direct comparison with the templates less realistic.
Even convolving the ACS templates with a Gaussian kernel to simulate the larger photometric errors in our sample would not be enough to reproduce the excess bGCs in our sample.
Therefore, while our aim remains to introduce this method as a possible avenue to study the evolution of the bright end of the galaxy luminosity function, more detailed studies are required to understand the potential of this method. This could be tackled, for example, in idealised simulations.
Still, using the HorizonAGN simulations, \cite{Brown24} have computed the galaxy mass function of galaxies that fell into clusters after $z \sim 2$, and found a slope of $\alpha = -1.20_{0.03}^{0.03}$. 
This value is much more constrained than the one obtained in Fig. \ref{fig:LF_constraint}, while still lying within the error bars.

\section{Summary}
\label{sec:conclusion}

We have carried out a {deep ($m_V\lesssim$ \SI{25.3}{mag}) photometric} study of GC candidates in the Hydra\,I cluster in the $U$, $V$, and $I$ bands.
Our $VI$ images extend out to $\sim$ \SI{265}{kpc} from the cluster centre, while the $U$-band data is restricted to the central region.
At the distance of Hydra\,I, we are able to probe $\sim$34\% of the GC luminosity function with a {sample size of 4886 GC} candidates.
We conclude that:

\begin{enumerate}
    \item The $V-I$ colours of our GC candidates are well described by a bimodal distribution. Blue GCs have a more extended distribution than red GCs, which are concentrated around the two massive central galaxies NGC\,3311 and NGC\,3309. The central distribution of bGCs are significantly displaced from the central galaxies, spatially coinciding with a secondary peak in X-ray emission and with a faint stellar envelope;
    \item The spatial distribution of very blue and very red GCs reveal an opposite displacement with respect to NGC\,3311, providing further evidence for the sloshing scenario of NGC\,3311;
    \item With $U$-band data, we are able to separate the central GCs in their ages and metallicities. We find that young metal-rich GCs are the most offset from the central galaxies, with overdensities coinciding with the ram-pressure tails of NGC\,3312 and with a group of dwarf galaxies to the northeast of NGC\,3311. Old metal-poor GCs are less displaced from the central galaxies, with overdensities to the south and north of NGC\,3311 that trace low surface brightness features. Finally, old metal-rich GCs seem to be associated with the central galaxies.
    \item The number density profiles of rGCs and bGCs show a striking difference: rGCs follow the surface brightness profile of NGC\,3311 out to large distances, while bGCs have a shallower profile. At distances {larger than \SI{100}{kpc},} the bGCs profile flattens and trace the intracluster light;
    \item The localised specific frequency ($S_N$) profile, calculated in annuli centred on NGC\,3311, shows that {rGCs maintain a flat profile with $S_N \sim 2.5$ for all radii, while bGCs display a sharp increase to $S_N \sim 15$ at $d\gtrsim$ \SI{100}{kpc}, further indicating a change of regime at this distance;}
    \item bGCs are better tracers of the total mass of Hydra\,I than rGCs;
    \item The localised $S_N$ profile may provide insights into the evolution of the faint-end slope of the galaxy luminosity function. Under the assumption that the intracluster component of GCs is dominated by bGCs and that these form mainly in low-mass galaxies, we mapped the $S_N$ values to typical galaxy magnitudes and calculated how many of those should have been destroyed to build up the light observed today.
    If these galaxies are added to the present-day luminosity function, {we derive a past faint-end slope of $\alpha=-1.81_{-0.16}^{+0.16}$ consistent with measurements in high-redshift galaxy clusters;}
    \item The colour distribution of GCs could be used to constrain the bright end of the luminosity function.
    We attempted to use a set of profiles of $V-I$ colour distributions from the Virgo cluster as templates to describe the colours in Hydra\,I, and found that Hydra\,I has an excess of GCs with very blue colours.
    This may be due to the different dynamical state of the two clusters.
    We sum the total contribution of the templates and obtain the number of galaxies necessary to produce the amount of GCs observed in Hydra\,I, from which we infer the past bright-end of the luminosity function.
    Fitting a Schechter function yields a {slope of $\alpha=-1.19_{-0.34}^{+0.21}$,} in agreement with cosmological simulations;
\end{enumerate}

Our results corroborate conclusions from previous works that the Hydra\,I cluster is in an active assembly stage.
The odd spatial distributions of red and blue GCs point to a sloshing motion of NGC\,3311, as suggested in \cite{Barbosa2018}, which will be the focus of a future work.
Moreover, the connection of the ICGC population with the former galaxy luminosity function outlined in our experiments is promising and may provide a new avenue to study its evolution, ultimately constraining galaxy and ICL formation models.
While our analysis presents several strong caveats, our goal was to introduce this new technique, because it can be applied to different clusters for which sufficiently deep imaging is available. A further validation of the method can be obtained by analysing cosmological simulations that include recipes for GC formation \citep[e.g.][]{Kruijssen19}.
On the observational side, the Euclid Wide Field Survey \citep{Scaramella22} promises excellent photometry down to $\mu_\textrm{VIS} \gtrsim 29.5$ \SI{}{mag/arcsec^2} for major local galaxy clusters in the coming years (e.g. Coma, Fornax, Virgo, and also Hydra I), enabling unprecedented low surface brightness studies and providing complete catalogues of GC systems, which can be used to further apply our methods.

\begin{acknowledgements}
      Based on observations collected at the European Organisation for Astronomical Research in the Southern Hemisphere under ESO programmes 65.N–0459(A), 80.A-0647(B) and 82.A-0894(A, B).
      FSL thanks the support from the IMPRS on Astrophysics at the LMU Munich.
      MS acknowledges the support of the Italian Ministry for Education, University and Research (MIUR) grant PRIN 2022 2022383WFT 'SUNRISE', CUP C53D23000850006, and by VST funds.
      MM acknowledges support from the ESO Studentship program.
      A.B. acknowledges support by the Deutsche Forschungsgemeinschaft (DFG, German Research Foundation) under Germany’s Excellence Strategy - EXC 2094 - 390783311.

\end{acknowledgements}

\bibliographystyle{aa} 
\bibliography{aa56513-25}

\begin{appendix} 
\FloatBarrier
\section{Selection of point sources}
\label{appendix:point_source_selection}

\begin{figure}
   \centering
   \includegraphics[width=\columnwidth]{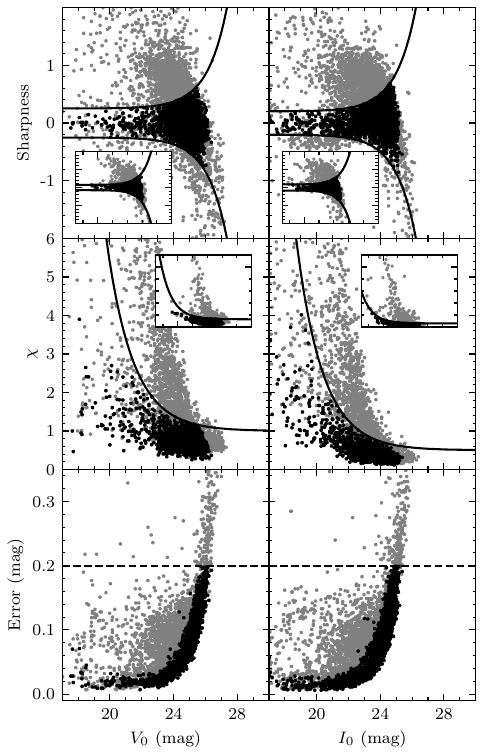}
   \caption{Point source selection from the photometry in the extinction-corrected $V$ and $I$ bands (left and right columns, respectively). The black exponential curves show the threshold used for our cuts in sharpness and $\chi$ (top and middle panels, respectively), while the dashed horizontal line indicates the adopted limit for the photometric error (bottom panel). The black points are the sources that satisfy our criteria in all three parameters, which we refer as point sources. For better visualisation, we only plot a randomly drawn subsample of 5000 data points. {The insets show the respective cuts for the central field, where our photometry is more accurate. The insets are plotted with the same axis ranges as in the main panel.}}
              \label{fig:point_source_selection_vi}
    \end{figure}

\begin{figure*}
\includegraphics[width=\textwidth]{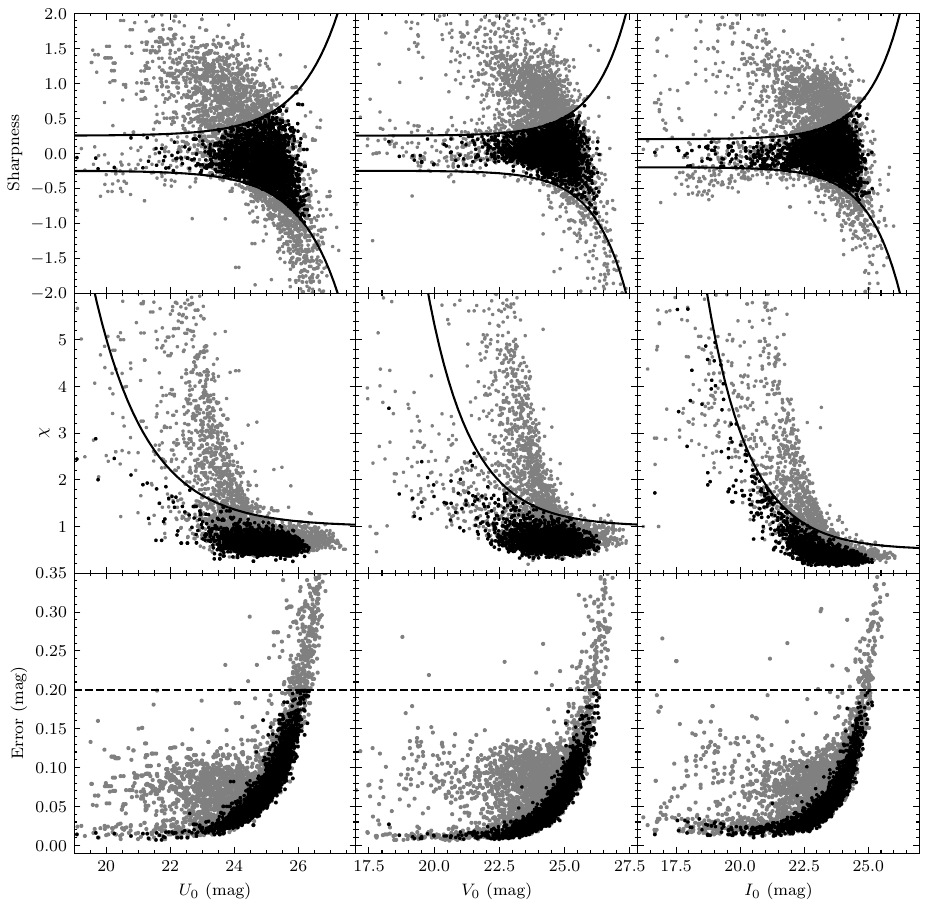}
\caption{Selection of point sources in our matched $UVI$ photometry.
As in Fig. \ref{fig:point_source_selection_vi}, grey points show all detected sources, while black points show those that satisfy the constraints in sharpness, $\chi$, and photometric error simultaneously.}
\label{fig:uvi_point_sources}
\end{figure*}

Figures \ref{fig:point_source_selection_vi} and \ref{fig:uvi_point_sources} show the selection criteria for point sources in the area covered by the $VI$ images and the $U$-band image, respectively.
This selection results in {9655} objects for the $VI$ sample, and 2729 objects for the $UVI$ sample.

\FloatBarrier
\section{Central completeness curves}
\label{appendix:completeness}

\begin{figure*}
    \centering
    \includegraphics[width=\textwidth]{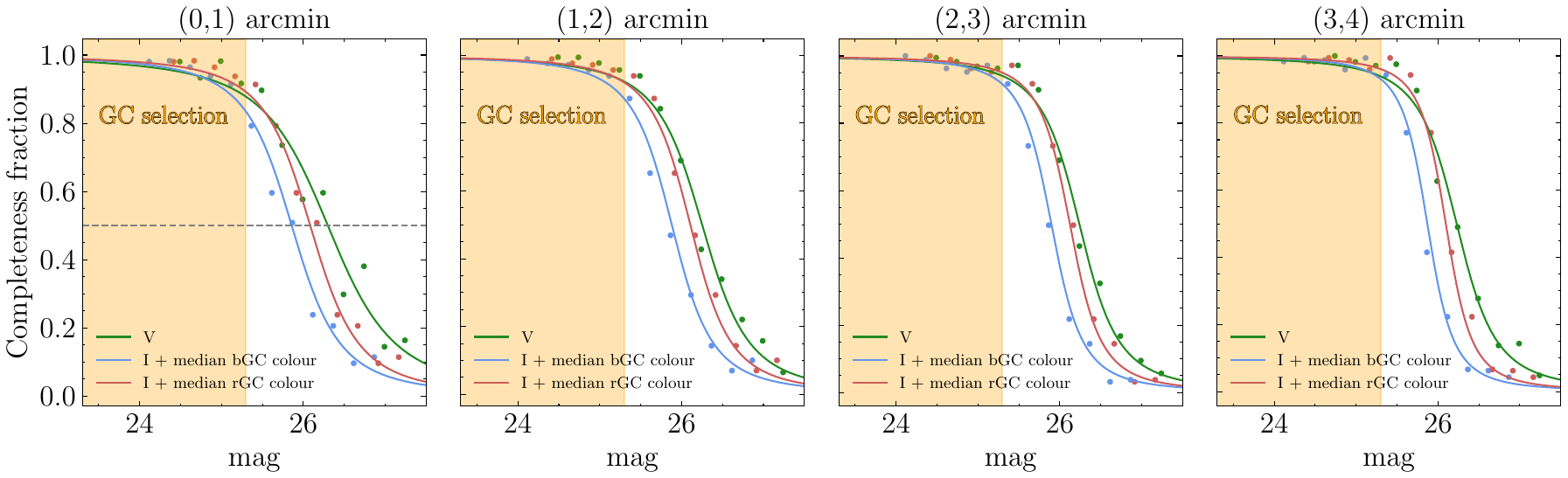}
    \caption{{Completeness fractions for different radial bins parametrised with a Pritchet function. The green curves show the $V$-band completeness, while red and blue curves show the $I$-band completeness shifted by the median colour of rGCs and bGCs, respectively. The orange shaded region shows the depth of our GCs candidate selection.}}
    \label{fig:completeness}
\end{figure*}

{The completeness curves shown in Fig. \ref{fig:completeness} are derived from artificial star injection--recovery experiments carried out as a function of both magnitude and radius. Artificial sources spanning the full magnitude range of the GC sample are randomly added to the galaxy-subtracted images following a uniform probability distribution in magnitude and position, to which we apply the same detection and PSF photometry as applied to the real data. The recovery fraction in each magnitude bin is computed separately in several radial bins, yielding completeness curves that quantify the detection efficiency as a function of radius.}

{Completeness curves are derived independently for the red and blue GC subpopulations. In the innermost radial bins, blue GCs are systematically less complete than red GCs at fixed magnitude, as illustrated in Fig. \ref{fig:completeness}. This difference is primarily driven by the shallower completeness in the $I$ band, which preferentially affects blue GCs due to their intrinsically bluer colours and therefore fainter $I$-band magnitudes at fixed $V$. At larger radii, where crowding is reduced and galaxy light residuals are lower, the completeness curves for red and blue GCs converge and the colour dependence becomes negligible.}

\section{Disruption of dwarf galaxies}
\label{appendix:tidal_disruption}

\begin{figure}
    \centering
    \includegraphics[width=\columnwidth]{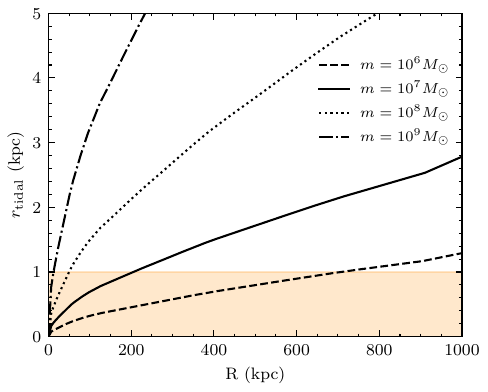}
    \caption{Tidal radius for galaxies of different stellar masses as a function of distance to the centre of Hydra\,I considering the enclosed mass derived from the X-ray emission.
    The orange shaded area indicates the regime where dwarf galaxies lose a significant amount of stars due to tidal stripping, delimited by the typical dwarf galaxy half-light radius of \SI{1}{kpc}.}
    \label{fig:tidal_rad}
\end{figure}

The scenario where metal-poor GCs come mainly from disrupted dwarf galaxies can be tested against theoretical predictions of the tidal stripping of these systems.
The tidal radius $r_\textrm{tidal}$ is the distance to an object of mass $m$ beyond which test masses are tidally disrupted by a primary, more massive body with mass $M$.
It is approximated to first order \citep{BinneyTremaine08} by

\begin{equation}
    r_\textrm{tidal} = R\, \left( \frac{m}{3M} \right)^{1/3}\, ,
\end{equation}

where R is the distance between the two bodies.
Figure \ref{fig:tidal_rad} shows the tidal radius as a function of projected distance for typical masses of dwarf galaxies, considering $M$ as the enclosed mass obtained from the X-ray model of \cite{Hayakawa06}.
The shaded region indicates $r_\textrm{tidal} \leq$ \SI{1}{kpc}, a typical value for the half-light radius of dwarf galaxies.
This marks the region where a given galaxy would lose at least half of its stars in a circular orbit, considering the potential observed in the present day.
Under these assumptions, less massive systems start losing most of their stars at large distances ($\sim$ \SI{200}{kpc} for $m=$ \SI{e7}{M_\odot}), while more massive ones sink deeper in the cluster before losing their stars to the intracluster space ($\sim$ \SI{50}{kpc} for $m=$ \SI{e8}{M_\odot}).

\FloatBarrier
\section{Globular cluster colour templates}
\label{appendix:templates}
\begin{figure}
    \centering
    \includegraphics[width=\columnwidth]{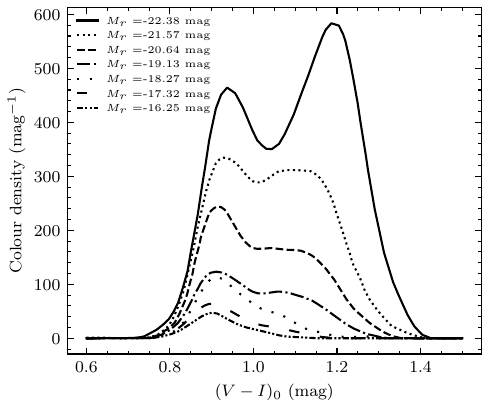}
    \caption{Colour distributions for different host galaxy magnitudes, adapted from \cite{Peng2006}. The templates are normalised by their specific frequencies. For better visualisation, the brightest and second brightest templates are rescaled by 1/3 and 2/3 of their original size, respectively.}
    \label{fig:colour_templates}
\end{figure}

Figure \ref{fig:colour_templates} shows the colour distribution of GCs in the Virgo cluster for different galaxy magnitudes from the ACS Virgo Survey \citep{Peng2006}.
The colours were transformed from the original $(g-z)_0$ to $(V-I)_0$ using $(V-I)_0 = 0.489 + 0.447(g-z)_0 + 0.027(g-z)_0^2$ obtained from single stellar population models from \cite{BruzualCharlot03}, and the profiles were normalised by their specific frequency, such that integrating each profile yields the expected number of GCs for their respective magnitude. 
Additionally, the average magnitude of each profile was transformed from the original $B$-band to $r$-band using the more recent Extended Virgo Cluster Catalogue \citep{Kim14}.
The range of validity of this colour transformation is $0.77 < (g-z)_0 < 1.90$ \SI{}{mag}, and we note that the colour templates reach colours slightly below the lower limit.

\FloatBarrier
\section{Faint-end constraint excluding very blue GCs}
\label{appendix:faint_end_test}

\begin{figure}
    \centering
    \includegraphics[width=\columnwidth]{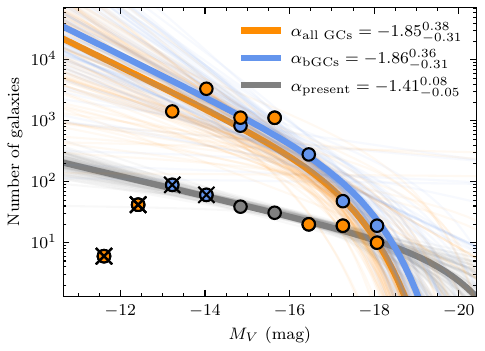}
    \caption{Evolution of the faint end of the galaxy LF excluding GCs with $(V - I)_0 < 0.8$ \SI{}{mag}. Colours and symbols are as in the top panel of Fig. \ref{fig:LF_constraint}.}
    \label{fig:faint_end_test}
\end{figure}

Since it is not possible to separate young from old GCs in the $VI$ sample, we perform a simple test by repeating the faint-end constraining experiment with a sharp cut in $V-I$ colour, including only GCs with $(V-I)_0>0.8$ \SI{}{mag}.
Figure \ref{fig:faint_end_test} shows the results under this condition.
As expected, by dropping the very blue GCs, we have much less galaxies contributing to the very faint part of the LF ($M_r \gtrsim -14$ \SI{}{mag} considering all GCs, and $M_r \gtrsim -16$ \SI{}{mag} considering only bGCs).
{This leaves us with less data points to fit the Schechter function, leading to larger error bars in the fit.}
We conclude that the very blue GCs are essential to predict a steeper slope by probing the progenitors at very faint magnitudes.

\FloatBarrier
\section{Faint-end constraint with an almost constant $S_N$ profile}
\label{appendix:faint_end_constant_sn}

T{he steepening of the $S_N$ profile is a natural result of the dynamical evolution of the cluster: dwarf galaxies (higher $S_N$) get disrupted at large distances and thus deposit their ‘high $S_N$ material’ at larger radii. In the method described in the main text, a constant $S_N$ profile would sample only a single galaxy with a characteristic $M_V$ corresponding to that particular $S_N$. By adding to the present LF, this would translate to a past LF identical to the present one, except for a peak in that particular $M_V$.}

\begin{figure}
    \centering
    \includegraphics[width=\columnwidth]{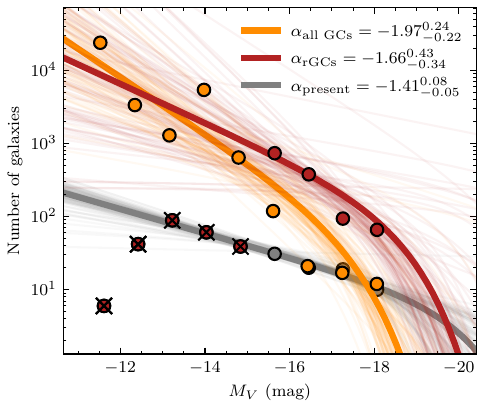}
    \caption{Evolution of the faint end of the galaxy LF considering the almost-constant $S_N$ profile of rGCs from Fig. \ref{fig:SN}. The corresponding past LF is shown with red markers.}
    \label{fig:faint_end_constant_sn}
\end{figure}

The idea of a single-valued $S_N$ profile for bGCs remains as a thought experiment and not really observed in real systems. Perhaps a more realistic (albeit unlikely) scenario for bGCs would be that of a non-increasing profile that fluctuates around a constant value, such as what is observed for rGCs in Fig. \ref{fig:SN}. The profile is not exactly flat, but revolves around $S_N \sim 2.5$. In Fig. \ref{fig:faint_end_constant_sn}, we apply our technique to such a profile and obtain a ‘bump’ at brighter magnitudes (in our case $M_V <$ \SI{-15}{mag}). The derived LF cannot be properly fitted by a Schechter function, as shown in the figure.

\FloatBarrier

\section{Faint-end constraint considering the scatter in the $S_N$-$M_V$ relation and GC occupation fraction}
\label{appendix:faint_end_refined}

In our method for constraining the evolution of the faint end of the luminosity function presented in Sec. \ref{sec:faint-end-constr}, two assumptions were considered: that the U-shape relation connecting $S_N$ and $M_V$ has no scatter, and that all destroyed galaxies have GCs (i.e. the GC occupation fraction is $f_\textrm{GC}=1$ galaxies of all magnitudes).
The observed $S_N$-$M_V$ relation presented in Figure 6 in \cite{Georgiev2010} shows that the scatter in $M_V$ can be significant for a given $S_N$.
Regarding $f_\textrm{GC}$, the general trend is that low-mass galaxies have also lower GC occupation (see Fig. 15 in \citealt{Carlsten22}).
In our original approach, assuming $f_\textrm{GC}=1$ means we are not considering the contribution from dwarf galaxies without GCs, who also contribute to the light budget in each radial bin.

To understand how these assumptions impact our results, we experiment with a slightly modified method.
Now, instead of associating each $S_N$ value to a single galaxy progenitor, we map them to a progenitor LF.
Specifically, we consider these progenitor LFs to be Gaussian, centred on the $M_V$ obtained from Eq. \ref{eq:SN_MV}. 
The dispersion of the Gaussian is not easy to estimate because the $S_N$-$M_V$ plane shows a discontinuous behaviour near galaxies hosting only a few GCs (see e.g. Fig. 20 in \citealt{Marleau25}).
For this reason, we fix $\sigma=$ \SI{0.9}{mag} as an illustrative but reasonable value.
This Gaussian LF is then corrected by the inverse of the GC occupation fraction, $1/f_\textrm{GC}(M_V)$, to account for galaxies that are GC-free, resulting in a Gaussian skewed to lower magnitudes.
The profile for $f_\textrm{GC}(M_V)$ was obtained from \cite{Carlsten22} considering a constant mass-to-light ratio $M/L=2$.
Finally, this primordial LF is normalised to match the luminosity measured in the radial bin.
This is done for all bins and added together with the present-day LF to obtain the constraint on the past Schechter slope.

\begin{figure}
    \centering
    \includegraphics[width=\columnwidth]{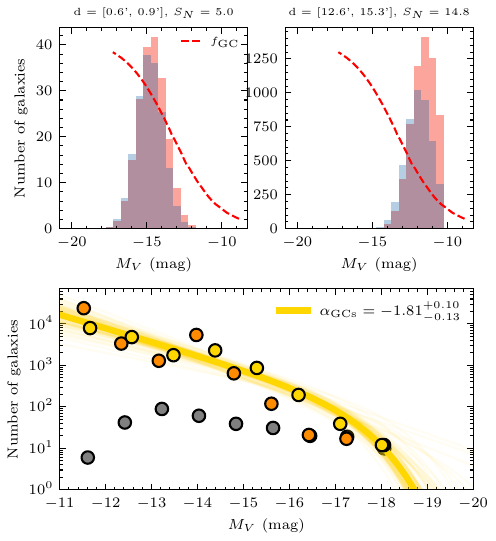}
    \caption{Constraint to the faint end of the past LF following the method described in Appendix \ref{appendix:faint_end_refined}. Top panels: examples of two radial bins and their associated primordial LFs according to the measured $S_N$. 
    In each panel, the blue histogram shows the Gaussian LF for the associated value of $S_N$, with dispersion $\sigma=$ \SI{0.9}{mag} and normalised to the luminosity inside the bin.
    The red histogram shows the skewed LF obtained after dividing the non-normalised Gaussian LF by the GC occupation fraction profile (red dashed line), and then normalising to the bin luminosity.
    Bottom: Faint-end constraint to the past LF considering the contribution from all annuli (gold), compared to the present-day LF (grey) and the past LF obtained with the simplified method (orange).}
    \label{fig:faint_LF_refined}
\end{figure}

Figure \ref{fig:faint_LF_refined} shows our results.
The two top panels show two examples of primordial LFs obtained in different annuli for their respective $S_N$.
As blue histograms, we show the primordial LFs considering $f_\textrm{GC}(M_V)=1$, while red histograms include the $1/f_\textrm{GC}(M_V)$ correction before normalising by the bin luminosity.
As expected, the GC occupation correction is stronger in the dwarf galaxy numbers, in particular for outer bins.
The lower panel of the figure shows the LF constraint obtained with this method.
{We predict a faint-end slope of $\alpha=-1.81_{-0.13}^{+0.10}$, which is still significantly steeper than the present-day value ($\alpha=-1.41_{-0.05}^{+0.08}$) and in excellent agreement with the values obtained with the simplified method.
Now, we get a larger contribution from brighter galaxies ($M_V\lesssim$ \SI{-16}{mag}), resulting from the dispersion of the individual primordial LFs.
We also verify that assuming a lower dispersion yields steeper LFs, while a larger dispersion results in a less constrained LF.}

\end{appendix}
\end{document}